\documentclass[11pt]{article}
\usepackage[body={17cm,23cm}]{geometry}

\usepackage[hang,small,center]{caption}
\usepackage{amsmath,amssymb}
\usepackage{amsfonts}
\usepackage{mathrsfs}
\usepackage{epic}
\usepackage{eepic}
\usepackage{graphicx}
\usepackage{cite}
\usepackage{enumerate}
\usepackage[curve]{xy}
\usepackage{xcolor}
\newcommand{\ds}{\displaystyle}

\renewcommand{\author}[1]{\large\rm #1\\ \bigskip}
\newcommand{\address}[1]{{\normalsize\it #1\\}\bigskip}
\renewcommand{\title}[1]{\bigskip\bigskip\Large\bf #1\bigskip\bigskip\\}

\newcommand{\Bigpsi}[3]{\phantom{\Psi}_2 \kern -.05em
\Psi_2\left(\genfrac{}{}{0pt}{}{#1}{#2}\biggl|#3\right)}

\newcommand{\bea}{\begin{eqnarray}}
\newcommand{\eea}{\end{eqnarray}}

\newcommand{\beq}{\begin{equation}}
\newcommand{\eeq}{\end{equation}}

\newcommand{\ii}{\mathsf{i}}

\newcommand{\cpar}{{\eta}}
\newcommand{\zedge}{\kappa_e}
\newcommand{\zsite}{\kappa_s}

\newcommand{\hf}{\frac{1}{2}}

\newcommand{\be}{\stackrel{\textrm{def}}{=}}
\newcommand{\crs}{\lambda}

\newcommand{\Gfun}{\Psi}
\newcommand{\G}{{\mathscr G}}
\newcommand{\Lscr}{{\mathscr L}}

\def\EXP{\textrm{{\large e}}}
\def\vt{\vartheta}
\def\re{\mathop{\hbox{\rm Re}}\nolimits}
\def\im{\mathop{\hbox{\rm Im}}\nolimits}

\def\tcpar{\widetilde{\cpar}}
\def\Lcal{{\mathcal L}}
\def\p{\phi}
\def\p{x}
\def\t{\alpha}
\def\ocpar{{\cpar}_0}

\newcounter{app}
\newcounter{sapp}[app]
\def\theapp{\Alph{app}}
\newcommand{\app}[1]{
\refstepcounter{app}{\vspace{7mm}
\noindent\Large\bf Appendix
\theapp.
 \ #1 \par \vspace{5mm}}
\setcounter{equation}{0}
\def\theequation{\Alph{app}.\arabic{equation}}}

\renewcommand{\theequation}{{\thesection}.{\arabic{equation}}}
\def\nsection#1{\setcounter{equation}{0}\section{#1}}

\begin{document}

\vglue 2cm

\begin{center}

\title{Quasi-classical expansion of the star-triangle relation
and integrable systems on quad-graphs}
\author{Vladimir V.~Bazhanov$^{1}$, \ Andrew P.~Kels$^{1,2}$ and \ Sergey M.~Sergeev$^{1,3}$}
\address{$^1$Department of Theoretical Physics, Research School of Physics and Engineering, \\
Australian National University, Canberra, ACT 0200, Australia}
\address{$^2$Institut f\"{u}r Mathematik, MA 8-4, Technische Universit\"{a}t Berlin,\\
 Str. des 17. Juni 136, 10623 Berlin, Germany.}
\address{$^3$Faculty of Education Science Technology \& Mathematics,\\
University of Canberra, Bruce ACT 2601, Australia.}
\end{center}

\begin{center}
{\sl Dedicated to Rodney Baxter on the occasion of his 75th birthday}
\end{center}
\vspace{1cm}

\begin{abstract}
In this paper we give an overview of exactly solved 
{\em edge-interaction} models, where
the spins are placed on sites of a planar lattice and interact through
edges connecting the sites. We only consider the case of
a single spin degree of freedom at each site of the lattice.
The Yang-Baxter equation for such models takes 
a particular simple form called the {\em star-triangle relation}. 
Interestigly all known solutions of this relation can be obtained 
as particular cases of a single ``master solution'', which is
expressed through the elliptic gamma function and have continuous
spins  
taking values on the circle. We show that in the low-temperature (or
quasi-classical) limit these lattice models reproduce classical discrete
integrable systems on planar graphs previously obtained and classified
by Adler, Bobenko and Suris through the 
{\em consistency-around-a-cube} approach. We also discuss inversion
relations, the physicical meaning of Baxter's
rapidity-independent parameter in the star-triangle relations and the
invariance of the action of the classical systems under the
star-triangle (or cube-flip) transformation of the lattice, which is a
direct consequence of Baxter's $Z$-invariance in the associated 
lattice models. 

\end{abstract}


\newpage
\setcounter{tocdepth}{2}
\tableofcontents
\nsection{Introduction}

The Yang-Baxter equation plays an exceptional role in statistical
mechanics and quantum field theory.
In particular, there exist many integrable models of statistical
mechanics constructed from solutions of the
Yang-Baxter equation. These models describe the interactions of discrete or
continuous spins (or fields) arranged on a two-dimensional lattice.
Here we only consider {\em edge-interaction} models, where
the spins are placed on sites of the lattice and interact through
edges connecting the sites. 
The Yang-Baxter equation for such models
takes its simplest ``star-triangular'' form \cite{Bax82}.
The most important models 
in this class include the Kashiwara-Miwa \cite{Kashiwara:1986}
and chiral Potts \cite{vG85,AuY87,Baxter:1987eq} models where spins 
take a finite set of integer values (both models also
contain the Ising model \cite{O44} and Fateev-Zamolodchikov
$Z_N$-model \cite{FZ82} as particular cases).  There are also
important continuous spin models, including the Zamolodchikov
``fishing-net'' model \cite{Zam-fish}, which describes certain planar
Feynman diagrams in quantum field theory, and the Faddeev-Volkov model
\cite{FV95}, connected with the quantization \cite{BMS07a} of discrete
conformal transformations \cite{Steph:2005}. 

Interestingly, all these
models, describing rather different physical problems, can be
obtained from particular cases of a single {\em master solution} of
the star-triangle relation found in \cite{Bazhanov:2010kz}. 
The master solution involves Boltzmann weights parameterized through the
elliptic-gamma function
\cite{Spiridonov1,Spiridonov4} 
and continuous spins taking values on a circle.  
It is worth noting that 
all the solutions mentioned above 
possess a {\em positivity property} and, therefore, can be
directly interpreted as local Boltzmann weights for physical solvable edge
interaction models on arbitrary planar graph.
In this paper we review main properties of these models
and establish their connection to classical discrete
integrable evolution equations, previously obtained and classified 
by Adler, Bobenko and Suris
(ABS) \cite{AdlerBobenkoSuris}. 

Before going into details of this correspondence it is useful to refer
to other recent remarkable appearances of the star-triangle relation
(and Yang-Baxter equation, in general) in different and seemingly
unrelated areas of physics and mathematics. In particular, 
there are deep connections to the theory of elliptic hypergeometric
functions \cite{Spiridonov2,Spiridonov3,Gahramanov:2013}, 
topological quantum field theory
\cite{Kashaev:2012cz,Andersen:2013rxa,Kashaev:2015} and calculations of  
superconformal indices connected with 
electric-magnetic dualities in 4D ${\cal N}=1$ superconformal
Yang-Mills theories \cite{Spiridonov:2011hf}. Indeed,
as found in \cite{Yamazaki:2012cp,  Terashima:2012cx,
  Xie:2012mr, Yagi:2015lha}, the 
4D superconformal quiver gauge theories are closely related to  
previously known 2D lattice models \cite{Bazhanov:2010kz, 
  Bazhanov:2011mz} and also lead to rather non-trivial new ones
\cite{Yamazaki:2013nra,Yamazaki:2015voa,Kels:2015bda,Gahramanov:2015cva}.
More generally, new advances were achieved in understanding the algebraic
structure of solutions of the Yang-Baxter equation, see, e.g., 
\cite{Chicherin:2012yn,Chicherin:2012jz,Derkachov:2013cqa,
  Chicherin:2014dya,Chicherin:2014fqa,Chicherin:2015mfv,
  Bazhanov:2010jq,Frassek:2011aa,Hernandez:2011,Bazhanov:2012ec,
Frenkel:2013uda,
  Bazhanov:2015gra,Meneghelli:2015sra}. Moreover, the 3D approach
\cite{Bazhanov:2005as,Mangazeev:2013spa} 
to quantum $R$-matrices resulted in an extremely concise expression 
\cite{Mangazeev:2014gwa,Mangazeev:2014bqa} of
the higher-spin $R$-matrix of the $XXZ$ model in terms of $q$-Hahn
polynomials. Soon after it was shown \cite{Borodin:2016} that 
this $R$-matrix happens to satisfy a 
stochastisity property and defines local transition probabilities of
the most general integrable totally asymmetric simple exclusion
process (TASEP) on a line. 


The key to obtaining the classical integrable equations from the
lattice spin models
lies in the low-temperature (or quasi-classical) 
limit. Indeed, all our lattice 
spin models contain a parameter, which
can be identified with the temperature (in the
language of statistical mechanics) or with the Planck constant 
(in the language of
Euclidean quantum field theory). We denote this parameter by the
letter $\hbar$.    
Consider a general nearest neighbour edge-interaction
model defined on a planar graph ${\mathscr G}$. Denote its set of 
sites (vertices) as $V({\mathscr G})$ and 
the set of edge as $E({\mathscr G})$.  
  In the quasi-classical
limit, $\hbar\to0$, the appropriately scaled spin variables
$\{x_i\}$, $i\in V({\mathscr G})$ always become continuous.  The
leading asymptotics of the partition function (for fixed boundary conditions)
\beq\label{zqcl}
 Z=\exp\Big({-\frac{{\mathcal
       A}(x^{(cl)})}{\hbar}}\Big)\big(1+O(\hbar)\big),\qquad
\hbar\to0,
\eeq 
is expressed in terms of an action ${\mathcal
  A}(x)$ evaluated on the solution of the classical equation of motion 
\beq\label{cem}
\frac{\partial {\mathcal A}(x)}{\partial
  x_i}\Big\vert_{x_i=x_i^{(cl)}}=0,\qquad
i\in V({\mathscr
  G})\ .
\eeq
The action is defined as a sum over edges of the lattice 
\beq
{\mathcal A}(\p)=\sum_{(ij)\in E(\mathscr G)}
{\mathcal L}_{ij}(\p_i,\p_j)\,,
\eeq
where the function ${\mathcal L}_{ij}(\p_i,\p_j)$,
which can be identified with the Lagrangian
density of the associated classical system, is determined 
by the leading order quasi-classical asymptotics of the edge
Boltzmann weight.
For each site $i\in V(\mathscr{G})$ Eq.\eqref{cem}  gives a constraint that
involves only the spin $x_i$, and each of the nearest neighbour spins
connected by an edge to site $i$.  These equations give a rather
general form of the so-called {\it discrete Laplace system} of
equations on $\mathscr{G}$ \cite{BS09}. Note, that the positivity
property of the 
underlying lattice models in many cases automatically leads to a
convexity property for the variational equations \eqref{cem}. From the
classical theory side convex variational priciples for the ABS
equations were studied in \cite{BG12}.

For lattice models the star-triangle relation may be considered as an
equation that  
connects the partition function of a 3-edge ``star'' graph, consisting
of one internal spin $x_0$ connected to three boundary spins
$x_1,x_2,x_3$, and a 3-edge ``triangle'' graph with vertices
$x_1,x_2,x_3$.   Since this relation defines an equality, in the
leading order quasi-classical expansion \eqref{zqcl} one 
obtains a {\it classical star-triangle relation} \cite{BMS07a} 
\beq\label{cstr0}
\mathcal{A}_\bigstar (\p_0^{(cl)},\p_1,\p_2,\p_3) =
\mathcal{A}_\triangle(\p_1,\p_2,\p_3),
\eeq
which equates the actions of the star and triangle graphs.  
Here $\p_0^{(cl)}$ is the solution to the equation 
\begin{equation}\label{cem2}
\frac{\partial{\mathcal A}_\bigstar(x)}{\partial \p_0}\Big|_{x_0=x_0^{(cl)}}
=0\ .
\end{equation}
which is the only equation of motion \eqref{cem} arising in the case of the 
3-edge star graph.
This equation provides a constraint on the four spin variables
$x_0,x_1,x_2,x_3$. Its precise form is, of course, model-dependent. 
Quite remarkably, exactly the same constrants arise in 
the ABS classification of the classical discrete integrable
equations on quad-graphs. More precisely, Eq.\eqref{cem2} can be
interpreted \cite{BS09} as the so-called {\it three-leg form} of the discrete
evolution equation for an elementary quarilateral.
We show that for all known solutions of the star-triangle equation the
constraint equations \eqref{cem2} always reduces those appearing in
the ABS classification list.  


Another important property of the classical star-triangle relation
\eqref{cstr0}, is that this relation implies the invariance of the
action functional $\mathcal{A}(x)$, under ``star-triangle''
transformations of the lattice \cite{BMS07a,BS09,Bazhanov:2010kz},
which is a natural counterpart of Baxter's $Z$-invariance for lattice
models. The classical star-triangle relation \eqref{cstr0} and the
associated constraint equation \eqref{cem2} are also related
\cite{BS09}  
to the 3D consistency relation \cite{AdlerBobenkoSuris} and more
generally to the multiform Lagrangian structures
and multidimensional consistency \cite{BS09,LNQ09}, further studied in
the recent papers 
\cite{Ball1108,Ball1302,Suris1307,Ball1307,Bobenko1403,Suris1510}.

This paper is structured as follows.  Section \ref{str-sect} gives an
overview of integrable models of statistical mechanics and their
properties, including details of the inversion relations for edge interaction
models.
Section \ref{sec:qcl} describes how one obtains discrete
integrable equations in general, by considering the quasi-classical
limit of an edge interaction model whose Boltzmann weights possess the
crossing symmetry. Special consideration are given to understanding
physical regimes of the resulting equations.  
Section \ref{sec:models} contains a review of all
exactly solved edge interaction models, including explicit definitions
of the Boltzmann weights, the partition function per edge (in the
thermodynamic limit), the corresponding classical Lagragian and its
relation to the ABS list \cite{AdlerBobenkoSuris} 
of integrable quad equations. Details of calculations are presented in
six Appendices. The main results of the paper are briefly summarized in
Conclusion.


\nsection{Star-triangle relation}\label{str-sect}
This introductory section summarizes important facts about the
star-triangle relation and solvable edge-interaction models on general
planar graphs. It contains a brief review of relevant results of
\cite{Bax1,Bax2,Bax02rip,BMS07a}, as well as some new additions to
the inversion relation technique \cite{Str79, Zam79, Bax82inv}.
\subsection{Edge-interaction models}
\begin{figure}[bt]
\centering
\includegraphics[scale=0.5]{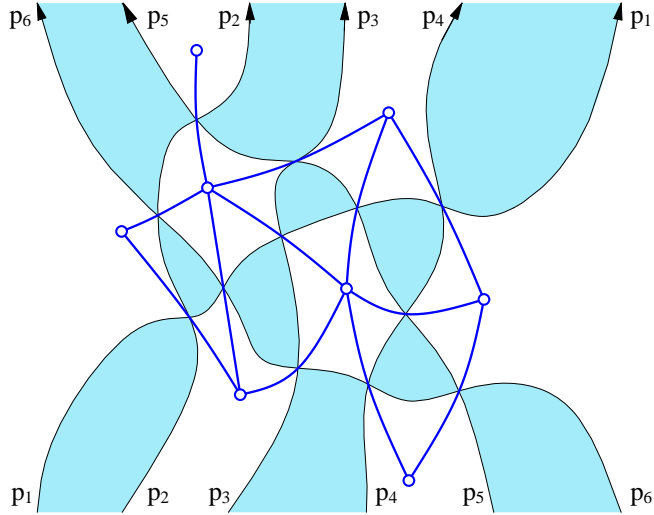}
\caption{The planar graph $\mathscr{G}$ (shown by open circles and bold edges)
and its medial graph $\Lscr$ (shown by thin edges and alternatively
shaded faces). }
\label{fig-net2}
\end{figure}
A general solvable edge-interaction model on a
planar graph can be defined in the following way \cite{Bax1,Bax2}.
Consider a planar graph $\G$, of the type shown in Figure \ref{fig-net2},
where its sites (or vertices) are drawn by open circles and the edges
by bold lines.
The same figure also contains another graph $\Lscr$, shown by thin lines,
which is the {\em medial graph} for $\G$. The faces of ${\Lscr}$ are
shaded alternatively; the sites of $\G$ are placed on the unshaded
faces.
We assume that for each line of $\Lscr$ one can assign a direction,
so that all the lines head generally from
the bottom of the graph to the top.
They can go locally downwards, but there can be no
closed directed paths in $\mathscr{L}$. This means that one can
always distort $\mathscr{L}$, without changing its topology, so
that the lines always head upwards\footnote{%
This assumption puts some restrictions on
the topology of the planar graph $\G$, but still allows enough generality for
our considerations here.}.
For further reference, let $F({\mathscr G})$, $E({\mathscr G})$
and $V({\mathscr G})$ denote respectively the
set of faces, edges and sites (vertices) of ${\mathscr G}$, and
$V_{int}({\mathscr G})$ the set of interior
sites of ${\mathscr G}$. The latter correspond to
interior faces of $\mathscr{L}$ (with a closed boundary).

Now we define a statistical mechanical spin model on $\mathscr{G}$.
To each line $\ell$ of ${\mathscr L}$
associate its own ``rapidity'' variable $p_{\ell}$, taking real values.
At each site $i$ of $\mathscr{G}$ place a spin $\sigma_i$, which take values
in some set. For the purpose of this
introduction, it is convenient to assume that
the spins are discrete and take a finite number $N\ge2$ of different values
$\sigma_i=0,1,\ldots,N-1$. In the following sections we will also consider
discrete spins, taking arbitrary integer values $\sigma_i\in{\mathbb Z}$ and
continuous spins, taking arbitrary values
$\sigma_i\in {\mathbb  R}$ on the real line.

Two spins interact only if they are connected by an edge.  This means
that each edge is assigned a Boltzmann weight that depends only on
spins at the ends of the edge. Usually, this weight depends on some
global parameters of the model (for instance, the temperature-like variables),
which are the same for all edges. Here we consider the case when the edge
weight also depends on a local parameter, given by the difference of
two rapidity variables associated with the edge.
\begin{figure}[hbt]
\begin{center}
\setlength{\unitlength}{.17in} \thicklines
\def\punit#1{\hspace{#1\unitlength}}
\def\pvunit#1{\vspace{#1\unitlength}}
\begin{picture}(4,9)\put(0,1)
{\begin{picture}(4,8)(0,-4.0)
\put(-6.5,0){\line(1,0){5.1}} \put(-3.4,0){\line(1,0){1.0}}
\put(-6.5,-3.2){\makebox(0,0)[b]{\small \mbox{$p$}}}
\put(-1.5,-3.2){\makebox(0,0)[b]{\small \mbox{$q$}}}
\put(-0.4,-0.2){\makebox(0,0)[b]{\small \mbox{$b$}}}
\put(-7.6,-0.2){\makebox(0,0)[b]{\small \mbox{$a$}}}
\put(-4.2,-5.6){\makebox(0,0)[b]{\small \mbox{$W(p-q\,|\,
a,b)$}}} \multiput(-6.4,-2.2)(0.2,0.2){22}{\makebox(0,0)[b]{
\mbox{${.}$}}} \multiput(-6.4,2.0)(0.2,-0.2){22}{\makebox(0,0)[b]{
\mbox{${.}$}}} \put(-6.8,0){\circle{.5}} \put(-1.2,0){\circle{.5}}
\put(8.0,-2.4){\line(0,1){4.6}} \put(8.0,0.3){\line(0,1){1.0}}
\put(5.5,-3.0){\makebox(0,0)[b]{\small \mbox{$p$}}}
\put(10.5,-3.0){\makebox(0,0)[b]{\small \mbox{$q$}}}
\put(8.0,3.0){\makebox(0,0)[b]{\small \mbox{$b$}}}
\put(7.4,-3.6){\makebox(0,0)[b]{\small \mbox{$a$}}}
\put(7.8,-5.6){\makebox(0,0)[b]{\small \mbox{$W(\cpar-
p+q\,|\,a,b)$}}}
\multiput(5.8,-2.2)(0.2,0.2){22}{\makebox(0,0)[b]{\mbox{${.}$}}}
\multiput(5.8,2.2)(0.2,-0.2){23}{\makebox(0,0)[b]{\mbox{${.}$}}}
\put(8.0,-2.6){\circle{.5}} \put(8.0,2.4){\circle{.5}}
\put(-1.8,2.2){\vector(1,1){0.3}}
\put(-6.3,2.2){\vector(-1,1){0.3}}
\put(10.2,2.2){\vector(1,1){0.3}}
\put(5.9,2.2){\vector(-1,1){0.3}}
\end{picture}}\end{picture}\end{center}
\caption{Edges of the first (left) and second types and their
  Boltzmann weights.}\label{fig1}
\end{figure}
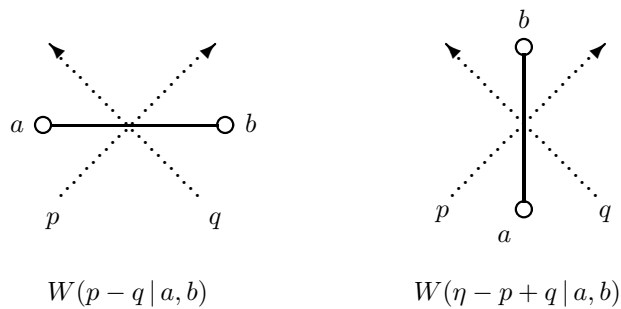
The detailed construction is as follows.
The edges of $\mathscr{G}$ are either of the first type in
Figure \ref{fig1},  or the second, depending on the arrangement of
the directed rapidity lines with respect to the edges.
For each edge introduce a ``rapidity
difference variable'' $\alpha_e$ defined as
\begin{equation}
\alpha_e=\left\{\begin{array}{ll}
p-q,\ \ \ \ \ \ \ \mbox{for an edge of the first type,}\\[.3cm]
\cpar-p+q,\ \mbox{for an edge of the second type,}
\end{array}\right.\label{difvar}
\end{equation}
where $p$ and $q$ are the rapidities, arranged as in Figure \ref{fig1}.
The constant $\cpar$ is a model-dependent parameter.
To each edge assign a Boltzmann weight factor $W(\alpha_e\,|\,a,b)$,
which depends on the rapiditity difference variable\footnote{%
The only known solvable edge-interaction model which does not have this
``difference property'' is the chiral Potts model
\cite{AuY87,BPA88}. Its quasi-classical limit has been studied in
\cite{Baz08} and will not be considered here.}
$\alpha_e$ and the spins $a$, $b$ at the ends of the edge.
Here we consider ``reflection-symmetric'' models\footnote{However there are also the Gamma function models with asymmetric Boltzmann weights, which are separately addressed in Section \ref{sec:gammamodels}.},

\beq
W(\alpha_e\,|\,a,b)=W(\alpha_e\,|\,b,a),\label{sym}
\eeq
where the weight $W(\alpha_e\,|\,a,b)$  is
unchanged by interchanging the spins $a$ and  $b$.
The property that the weight functions for the edges of two types in
Figure \ref{fig1} are obtained from each other by the
transformation $\t\to\cpar-\t$ of rapidity difference variable
$\alpha$ is called the {\em crossing symmetry}. By this reason the
parameter $\cpar$ in \eqref{difvar} is usually called the ``crossing
parameter''.

In general, there may also be a single-spin self-interaction with a
rapidity-independent weight $S(a)$ for each spin $a$.
The partition function is defined as
\begin{equation}
Z=\sum_{\{\sigma\}}  \ \prod_{i} S(\sigma_i)\
\prod_{(ij)}
W(\alpha_{(ij)}\,|\,\sigma_i,\sigma_j)\ ,\label{Z-def}
\end{equation}
where the first product is over all sites $i\in V({\mathscr
  G})$ and
the second is over all edges $(ij)\in
E({\mathscr G})$. The sum is taken over
all possible values of the {\em interior} spins (in the case of continuous
spins, the sum is replaced by an integral). The boundary spins are kept fixed.

\subsection{Star-triangle relation}\label{str-subsec}

\begin{figure}[hbt]
\begin{center}
\includegraphics[scale=0.45]{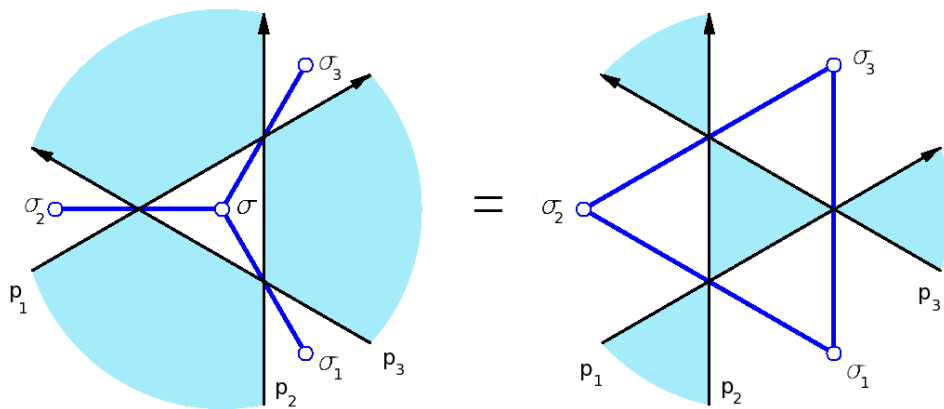}
\end{center}
\caption{A pictorial representation of the star-triangle relation
  \eqref{str-def}.}
\label{startriangle}
\end{figure}

Integrability of the model requires that the Boltzmann weights satisfy the
star-triangle relation \cite{Bax82}. For the reflection-symmetric
case \eqref{sym} this relation reads
\beq\begin{array}{l}
\ds\sum_{\sigma} S(\sigma) \,{W}(\cpar-\alpha_1\,|\,a,\sigma)
\,{W}(\cpar-\alpha_2\,|\,b,\sigma)\,W(\cpar-\alpha_3\,|\,c,\sigma)\\[.4cm]
\phantom{\ds\sum_{\sigma} S(\sigma) {W}(\cpar-\alpha_1\,|\,a,\sigma)
\,{W}(\cpar)}
={\cal R}_{123} \,W(\alpha_1\,|\,b,c)\,
W(\alpha_2\,|\,a,c)  \,
W(\alpha_3\,|\,a,b),
\end{array}
\label{str-def}
\eeq
where
\beq
\alpha_1+\alpha_2+\alpha_3=\cpar,
\eeq
and $ {\cal R}_{123}$ is some scalar factor independent of the spins
$a, b, c$.
For continuous spins the sum in \eqref{str-def} is replaced by an
integral.  The star-triangle relation equates partition functions of
the ``star'' and ``triangle'' graphs shown in Figure \ref{startriangle},
where the external spins $a$, $b$ and $c$ are fixed. The
variables $\alpha_1,\alpha_2,\alpha_3$ in \eqref{str-def} are
calculated, according to the rule \eqref{difvar}, from the rapidity
variables $p_1,p_2,p_3$ shown in Figure \ref{startriangle},
\beq
\alpha_1=p_2-p_3,\qquad \alpha_2=\cpar-p_1+p_3,\qquad
\alpha_3=p_1-p_2.\qquad
\eeq
Note that, there is also a second star-triangle relation
corresponding to a mirror image of Figure \ref{startriangle}.
However, for the reflection-symmetric
models this relation is equivalent to \eqref{str-def}.

\subsection{The factor  $ {\cal R}_{123}$ }\label{factor-R}

As shown in \cite{MatSmirn90,Bax02rip} the factor ${\cal R}_{123}$
appearing in \eqref{str-def} can be conveniently expressed in terms of the
weights $W(\alpha\,|\,a,b)$.
The result applies in the case of discrete spins.
It is based on a property that the quantity
\beq
P(\alpha)=\prod_{b} W(\alpha\,|\,a,b), \label{P-def}
\eeq
is spin-independent, {\it i.e.}, the same for all values of the
spin $a=0,1,\ldots,N-1$. It seems that this additional
requirement is not a simple corollary of the symmetry \eqref{sym} or
the star-triangle \eqref{str-def} relations
themselves,  but it is certainly true for all their currently known
discrete spin solutions, particularly for those
considered in this paper.
Define the function
\beq
f(\alpha)^N=
P(\alpha)^{-1}\
{\ds\det\Vert \,S(a) \,W(\cpar-\alpha)\,|\,a,b)\,\Vert_{0\le a,b\le
    N-1}} .\label{fpq-def}
\eeq
Now,
regard the spin $c$ as fixed and consider each side of
\eqref{str-def} as the element $(a,b)$ of some matrix. Taking the
determinant of this matrix one obtains
(to within an undetermined factor of an $N$-th root of unity),
\beq
{\mathcal R}_{123}= f(\alpha_1)\,f(\alpha_2)/f(\cpar-\alpha_3)\ .\label{R-one}
\eeq
Because of \eqref{P-def} the result is independent of the value of the
fixed spin $c$. Repeat the same calculations two more times, replacing
the spin $c$ with the spin $a$ or $b$. One obtains another two
expressions for ${\mathcal R}_{123}$, similar to \eqref{R-one}, but
with permuted indices $1,2,3$ in the RHS. All three
expressions are consistent only if the quantity
\beq
\zsite^2=f(\alpha)\, f(\cpar-\alpha) \label{zs-def}
\eeq
is independent of the variable $\alpha$. Taking this into account one
can write \eqref{R-one} in a symmetric form
\beq
{\cal R}_{123}=f(\alpha_1)\,f(\alpha_2)\,f(\alpha_3)/\zsite^2\ .\label{Rpqr}
\eeq
The results \eqref{zs-def} and \eqref{Rpqr}
hold for any solutions of the star-triangle relation, having
the property \eqref{P-def}.

The quantity \eqref{zs-def} is the ``rapidity
independent parameter'', defined by Baxter
in \cite{Bax02rip}. To be more precise, Baxter defined the
``invariant'' $I= \zsite^2/N$  (see Equation (13)
of \cite{Bax02rip}) and showed that $I=1$ for self-dual and/or critical
models. Here we give a different physical interpretation to the
quantity $\zsite$ by identifying it with a single-site factor in the
asymptotic expression for the partition function
in the large-lattice limit (see
Equation \eqref{factor} below). From \eqref{fpq-def} and \eqref{zs-def} it
is easy to see that $\kappa_s$ linearly depends on the normalization of the
site weights $S(a)$, but does not depend on the normalization of
the edge weights $W(\alpha\,|\,a,b)$. Indeed, it scales
linearly with the site weights,
\beq
S(a)\to \lambda S(a),\qquad \kappa_s\to \lambda \kappa_s,
\eeq
but remains unchanged if the edge weight
$W(\alpha\,|\,a,b)$ is multiplied by any factor
independent of the spins $a, b$.

\subsection{Inversion relations}\label{prop-section}

\begin{figure}[hbt]
\begin{center}
\includegraphics[scale=0.6]{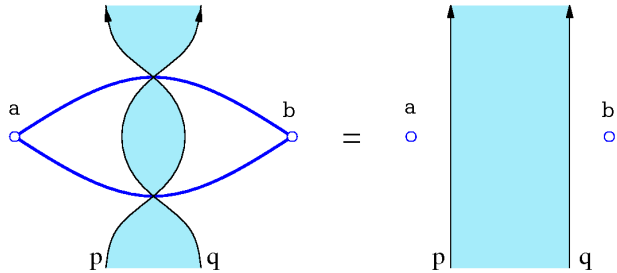}
\end{center}
\caption{A pictorial representation of the inversion relation \eqref{inv1}.}
\label{fig-i1}
\end{figure}

\begin{figure}[hbt]
\begin{center}
\includegraphics[scale=0.5]{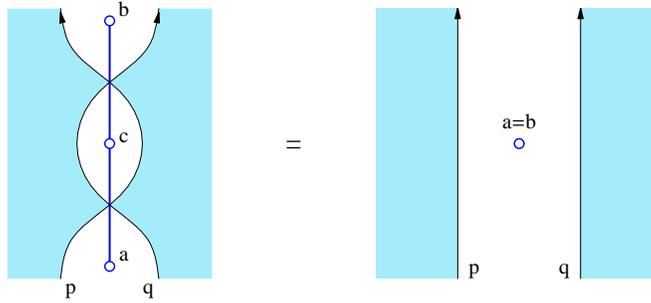}
\end{center}
\caption{A pictorial representation of the inversion relation \eqref{inv2}.}
\label{fig-i2}
\end{figure}

For all models considered here that satisfy the crossing symmetry,
the weights can be normalized so that they satisfy
\begin{enumerate}
\item[(a)] {\sl the boundary conditions}
\beq
W(\alpha\,|\,a,b)\big|_{\alpha=\cpar}=\delta_{ab}/S(a),
\qquad W(\alpha\,|\,a,b)\big|_{\alpha=0}=1,\qquad \forall
a,b,
\label{init}
\eeq
\end{enumerate}
Together with the star-triangle relations these conditions imply
\begin{enumerate}
\item[(b)] {\sl the inversion relations}
\bea
W(\alpha\,|\,a,b)\, W(-\alpha\,|\,a,b)&\!\!\!=\!\!\!&1,
\qquad \forall a,b, \label{inv1}\\[.4cm]
\ds S(a)\,\sum_c W(\cpar+\alpha\,|\,a,c)\,S(c)\,W(\cpar-\alpha\,|\,c,b)
&\!\!\!=\!\!\!&f(\alpha)\,f(-\alpha)\,\delta_{ab}\ .
\label{inv2}
\eea
\end{enumerate}
Note that \eqref{init} and \eqref{zs-def} imply
\beq
f(0)=1,\qquad f(\cpar)=\zsite^2.\label{f0}
\eeq
The two inversion relations \eqref{inv1} and \eqref{inv2} are
corollaries of the star-triangle relation \eqref{str-def} and boundary conditions \eqref{init}.
Indeed, comparing both sides of \eqref{str-def} with $\alpha_3=\cpar$ and
taking into account \eqref{init}, one concludes that the product
$W(\alpha_1\,|\,a,c)\, W(-\alpha_1\,|\,a,c)$
is independent of the spins $a,c$. Without
loss of generality this product could be normalized by the condition
\eqref{inv1}.
Substituting this condition back into \eqref{str-def}
(still keeping $\alpha_3=\cpar$) and using \eqref{Rpqr} and \eqref{f0},
one immediately obtains \eqref{inv2}.
The inversion relations \eqref{inv1} and \eqref{inv2} are
represented pictorially in Figure \ref{fig-i1} and Figure \ref{fig-i2}.
They
can be used to calculate the partition function \eqref{Z-def} in the
large-lattice limit.
For example, for a regular square lattice of $M$ sites
there are only two different rapidities $p$ and $q$. Correspondingly,
half of the edges have the rapidity difference variable \eqref{difvar} equal
to $\alpha=p-q$ and the other half to $\cpar-p+q=\cpar-\alpha$. Let
\beq
\kappa(\alpha)=Z^{1/M},\qquad M\to\infty,
\eeq
be the partition function
per site, then one can show that \cite{Str79, Zam79, Bax82inv},
\beq
\kappa(\alpha)\,\kappa(-\alpha)=f(\alpha)\,f(-\alpha),\qquad
\kappa(\alpha)=\kappa(\cpar-\alpha)\ . \label{func1}
\eeq
Together with an appropriate analyticity assumption (typically
$\log\kappa(\alpha)$ is analytic and bounded in a suitable domain
including the segment $0\le\alpha\le\cpar$) these equations uniquely
determine $\kappa(\alpha)$, see \cite{Str79, Zam79, Bax82inv} for
further details.

We have found that for all models with crossing symmetry considered here (and we believe
that it is a general property of solvable edge-interaction models),
the quantity $\kappa(\alpha)$ can be
factorized as
\beq
\kappa(\alpha)=\zsite\,\zedge(\alpha)\,\zedge(\cpar-\alpha),\label{kappa-sq}
\eeq
where $\zsite$ is defined in \eqref{zs-def} and the function $\zedge(\alpha)$
satisfies the functional equations
\beq
\zedge(\alpha) \,\zedge(-\alpha)=1,\qquad
\zedge(\cpar-\alpha)/\zedge(\alpha)=f(\alpha)/\zsite,\label{func2}
\eeq
\beq
\zedge(\cpar-\alpha)\,\zedge(\cpar+\alpha)=f(\alpha)\,f(-\alpha)/\zsite^2\ .
\label{func3}
\eeq
where the third equation is a corollary of the first two. It is easy
to see that if $\kappa_s$ and $\kappa_e(\alpha)$ satisfy \eqref{zs-def} and
\eqref{func2} then $\kappa(\alpha)$ solves \eqref{func1}.

There are exactly two edges (one of each type) for each site of a
regular square lattice. Correspondingly, the partition function per
site \eqref{kappa-sq} is a product
of the rapidity independent single-site
factor $\zsite$ and two single-edge factors $\zedge(\alpha)$ and
$\zedge(\cpar-\alpha)$ (see Equation \eqref{factor} below for the
generalization of \eqref{kappa-sq} for an arbitrary graph).

\subsection{Canonical normalization of the Boltzmann weights}
\label{canonical}
It is possible to further refine the normalization of the weights
to simplify the scalar factors in
the star-triangle relation \eqref{str-def}, and
in the second inversion relation \eqref{inv2}.
Indeed, it is easy to see that if one rescales the weight functions
\beq
W(\alpha\,|\,a,b)\to\mathcal{W}(\alpha\,|\,a,b)=\frac{1}{\zedge(\alpha)}\,W(\alpha\,|\,a,b),
\qquad S(a)\to\mathcal{S}(a)=\frac{1}{\zsite}\,S(a),\label{renorm}
\eeq
in the definitions \eqref{fpq-def}, \eqref{zs-def}, \eqref{Rpqr} and
\eqref{func2} then all the associated scalar factors become equal to one,
\beq
\widehat{\kappa}_s=1,\qquad
\widehat{\mathcal
  R}_{123}\equiv
\widehat{f}(\alpha)\equiv\widehat{\kappa}_e(\alpha)\equiv1.\label{rfactor}
\eeq
In what follows we will refer to this distiguished normalization as
the {\em canonical normalization of weights}.
For further reference, define a rescaled partition function
$\widehat{Z}$ obtained from \eqref{Z-def} by the substitution
\eqref{renorm},
\begin{equation}
\widehat{Z}=\sum_{\{\sigma\}}  \ \prod_{i} \mathcal{S}(\sigma_i)\
\prod_{(ij)}
\mathcal{W}(\alpha_{(ij)}\,|\,\sigma_i,\sigma_j)=Z\,\Big(
k_s^{M}\,
\prod_{(ij)}
\kappa_e(\alpha_{(ij)})\Big)^{-1}
 ,\label{Z-def1}
\end{equation}
where $M$ is the total number of sites of the graph ${\mathscr G}$ and the
rest of notations are the same as in \eqref{Z-def}.

\subsubsection{Positivity}
For most of the models considered below (except the asymmetric
models of Section \ref{sec:gammamodels}) the Boltzmann weights ${\mathcal S}(a)$
and $\mathcal{W}(\t\,|\,a,b)$ are real and positive when $\t$ is real
and lies in the physical domain $0<\t<\cpar$.

\subsection{$Z$-invariance}\label{Zinvar}
The partition function \eqref{Z-def1}
possesses remarkable invariance properties \cite{Bax1,VJones,Bax2}.
It remains unchanged (up to simple $f(\t_{ij}) $ and $\kappa_s$ factors)
by continuously deforming the lines of ${\mathscr L}$ with their boundary
positions kept fixed, as long as the graph ${\mathscr L}$
remains directed. In
particular, no closed directed paths are allowed to appear%
\footnote{Actually, these restrictions can be removed if one properly defines
``reflected'' rapidities for downward going lines (see Sect.3 of
\cite{Bax2}), but we will not elaborate this point here here.}%
.
It is easy to see that all such transformations
reduce to a  combination of the moves shown in
Figure \ref{startriangle}, Figure \ref{fig-i1}, and Figure \ref{fig-i2},
corresponding to the star-triangle \eqref{str-def} and inversion
relations \eqref{inv1}, \eqref{inv2}.
In general the partition function acquires simple  $f(\t_{ij}) $ and
$\kappa_s$ factors under these moves, however for the
canonical normalization \eqref{Z-def1} the invariance is strict (all
extra factors become equal to one in this case, see \eqref{rfactor}).
Given that the graphs ${\mathscr L}$ and
${\mathscr G}$ can undergo rather drastic
changes, the above ``$Z$-invariance'' statement is rather non-trivial.

The partition function \eqref{Z-def1} depends on the exterior spins
and the rapidity variables $p_1,p_2,$ $\ldots,$ $p_L$. Of course, it also
depends on the graph  $\mathscr L$, but only on a relative ordering
(permutation) of the rapidity lines at the boundaries and not on their
arrangement inside  the graph.
Naturally, this graph can be identified with an element of the
permutation group.
Then
the partition function $Z$ can be regarded as a group 
representation matrix, acting non-trivially on the spins at the lower and
upper boundaries (it acts as an identity on the leftmost and rightmost
spins in Figure \ref{fig-net2}, corresponding to unbounded faces).
Although the above $Z$-invariance holds for arbitrary values of rapidity
variables $p_1,p_2,\ldots,p_L$, it is convenient to distinguish
the {\em physical regime},
when all the rapidity differences $\alpha_{(ij)}$ lie in the
interval $0<\alpha_{(ij)}< \eta$. In this case all Boltzmann weights
entering \eqref{Z-def} and \eqref{Z-def1} are real and positive and
the corresponding partition functions have a straightforward
interpretation in statistical mechanics.
%
%
%
Note that for the physical regime the graph
$\mathscr L$ cannot contain more than one intersection for the same
pair of the rapidity lines.

Consider a generic graph $\mathscr{G}$ with
a large number of sites, $M$, and a large number of edges, $N\sim 2M$.
Then the number of boundary sites is on the order of $2\,M^{1/2}$.
Assume that the corresponding boundary spins are kept finite.
Then, following \cite{Bax1}, one can show
that the  leading asymptotics of the partition function
\eqref{Z-def1} at large $M$ has the form
\begin{equation}
\log Z= M\,\log{\kappa_s}+\sum_{(ij)\in E(\mathscr{G})}
\log\kappa_{e}(\alpha_{(ij)})+O(\sqrt{M})\,,\label{factor}
\end{equation}
where the factors $\kappa_s$ and $\kappa_e(\t)$ are defined in
\eqref{zs-def} and \eqref{func2}.  Note that the factors are universal;
they are independent of the graph ${\mathscr G}$.  This result holds for
any $Z$-invariant system with positive Boltzmann weights for a large
graph $\mathscr L$ with sites in general position. Evidently, for the
canonically normalized partition function \eqref{Z-def1}, the leading
term in \eqref{factor} (the bulk free energy) vanishes, 
\begin{equation}
\log{\widehat {Z}}=O(\sqrt{M}),\qquad M\to \infty\ .
\end{equation}

\begin{figure}[hbt]
\begin{center}
\includegraphics[scale=0.6]{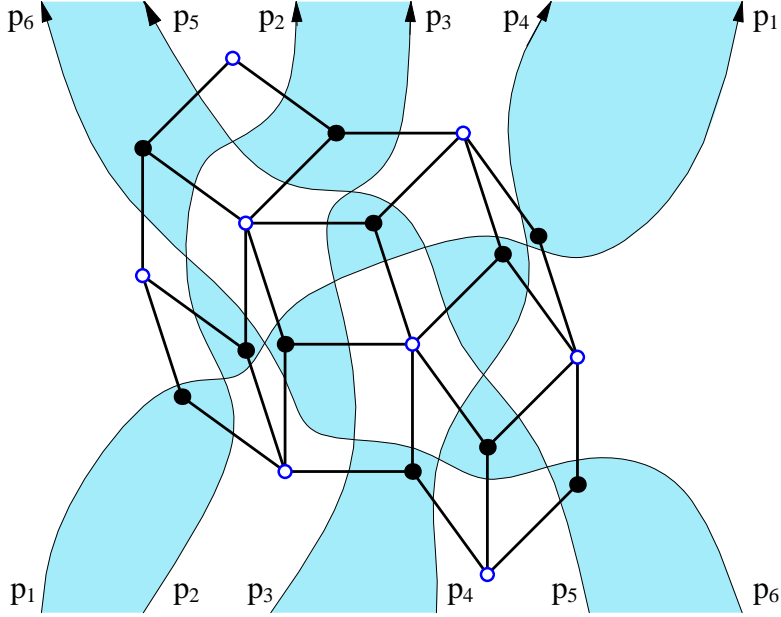}
\end{center}
\caption{A rhombic embedding of the graph ${\mathscr L}^*$}
\label{fig6}
\end{figure}

\subsection{Rapidity graphs and rhombic tilings}\label{tilings}

Consider some additional combinatorial and geometric structures
associated with the graph $\mathscr L$.
First, if the unshaded faces in the above definition of $\mathscr{G}$
are replaced by the shaded ones, one obtains another
graph $\mathscr{G}^*$, which is dual to $\mathscr{G}$.
Each site of $\mathscr{G}^*$ corresponds to a face of $\mathscr{G}$
and vice versa. Obviously, both graphs $\mathscr{G}$ and $\mathscr{G}^*$
have the same medial graph $\mathscr{L}$. Assign the difference
variables $\alpha_{e^*}$ to the edges of $\mathscr{G}^*$ by the same
rule \eqref{difvar}. Note that there is a one-to-one correspondence between the
edges of $\mathscr{G}^*$ and $\mathscr{G}$. Moreover, if $e\in
E(\mathscr{G})$ is of the first type then the corresponding edge $e^*\in
E(\mathscr{G}^*)$ is of the second (and vice versa).  In other words for the
corresponding edges $\alpha_e+\alpha_{e^*}=\cpar$.
Let ${star}(i)$ denote the set of edges meeting at the site
$i$. It is easy to show that for any interior site of $\mathscr{G}$
\begin{equation}
\sum_{(ij)\in {star}(i)} \alpha_{(ij)}=2\cpar,\qquad
i\in V_{int}(\mathscr{G})\ .\label{sumrule1}
\end{equation}
Indeed, consider a face of ${\mathscr L}$, containing the site $i$. 
It is bounded by the directed thin lines forming the graph ${\mathscr
  L}$, see Figure \ref{fig-net2}. Vertices of this face correspond to the edges
of ${\mathscr G}$ that meet at the site $i$. By construction, the lines of
${\mathscr L}$ are always heading upwards, so there must be exactly one lowest
and one highest vertex for each face of ${\mathscr L}$. These two
vertices correspond to the second type edges in Figure \ref{fig1}.
The remaining vertices correspond to edges of the first type. Taking
this into account, one immediately arrives to \eqref{sumrule1}.  
A similar sum rule holds for the dual graph ${\mathscr G}^*$,
\begin{equation}
\sum_{(kl)\in {star}(k)} \alpha^*_{(kl)}=2\cpar, \qquad
k\in V_{int}(\mathscr{G}^*)\ .\label{sumrule2}
\end{equation}

\begin{figure}[hbt]
\begin{center}
\includegraphics[scale=0.6]{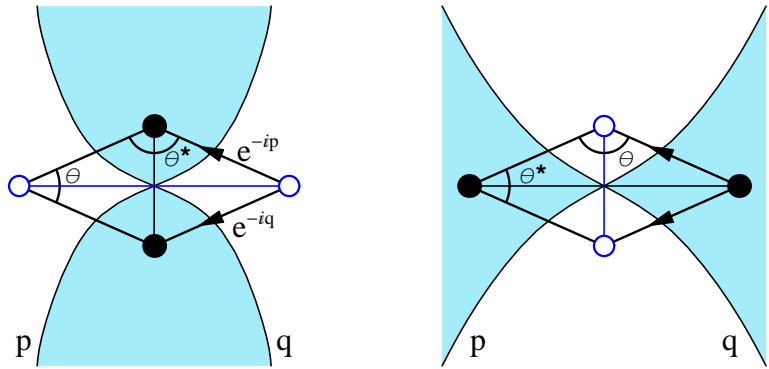}
\end{center}
\caption{Two types of rhombi.}
\label{two-rhombi}
\end{figure}

Consider yet another graph $\mathscr{L}^*$, dual to $\mathscr{L}$.
The set of sites of $\mathscr{L}^*$ consists of those
of the graph $\mathscr{G}$ and of its dual $\mathscr{G}^*$.
These sites are shown in Figure \ref{fig6} by white and black dots,
respectively.
The edges of ${\mathscr L}^*$ always connect
one white and one black site.
%
The faces of $\mathscr{L}^*$ correspond to the vertices of $\mathscr{L}$.
The latter are of degree four, therefore ${\mathscr L}^*$
is a ``quad-graph'' (a planar graph with quadrilateral faces). 
The edges of $\mathscr{G}$ and $\mathscr{G}^*$ are diagonals of these
quadrilaterals (see Figure \ref{two-rhombi}). Evidently, there are exactly two 
white and two black vertices in each face.
Remarkably, the graph $\mathscr{L}^*$ admits a {\em rhombic embedding}
into the plane.  In other words this graph can be drawn so that all
its edges are line segments of the same length and, hence, all of its
faces are rhombi, as shown in Figure \ref{fig6}.  The corresponding
theorem \cite{Kenyon} states that such an embedding exists if and only if
(a) no two lines of $\mathscr L$ cross more than once\footnote{%
The rapidity lines forming the graph $\mathscr L$ are called ``train
tracks'' in \cite{Kenyon}.}  and (b) no line of $\mathscr
L$ crosses itself or is periodic.
Note, that in the physical regime these conditions are obviously satisfied.

Assume that the edges of the quadrilaterals are of unit length, and consider
them as vectors in the complex plane.
To precisely specify a rhombic embedding one needs to provide the
angles between these vectors.
A rapidity line always crosses opposite (equal)
edges of a rhombus.
Therefore, all edges crossed by the same rapidity
line $p$ are given
by one vector $\zeta_p$,\  $|\zeta_p|=1$.
Thus, if the original rapidity graph ${\mathscr L}$
has $L$ lines, there will be only $L$
different edge vectors.
Choose them as $\zeta_{p_k}=e^{-i\pi p_k/\cpar}$, where
$p_1,p_2,\ldots,p_L$ are the corresponding rapidity variables.
Each face of $\mathscr{L}^*$ is crossed by exactly two rapidity lines
$p_k$ and $p_\ell$.
To this face associate a rhombus with the edges
$\zeta_{p_k}$ and $\zeta_{p_\ell}$, as shown in Figure \ref{two-rhombi}.
Its diagonals are
edges of $\mathscr{G}$ and $\mathscr{G}^*$.
The rhombus angles are
precisely the (rescaled) difference variables $\pi\alpha_e/\cpar$ and 
$\pi\alpha_{e^*}/\cpar$
assigned to these edges (this is true for both types of rhombi shown
in Figure \ref{two-rhombi}). In the physical regime all these angles are
in the range from $0$ to $\pi$. So the rhombi will have positive area
and will not overlap. The sum rules \eqref{sumrule1} and
\eqref{sumrule2} guarantee that the resulting ``rhombic tiling'' 
is flat with no cusps at the sites of ${\mathscr L}^*$. 
\nsection{Quasi-classical expansion} \label{sec:qcl}

\subsection{Boltzmann weights, partition function, Lagrangian and action}
The models considered below contain a free parameter, which can be
identified with the Planck constant (in the language of Euclidean
quantum field theory) or with the temperature (in the language of
statistical mechanics). We denote this parameter by the letter
$\hbar$. Its precise definition is model-dependent (see below), but it
is always possible to choose $\hbar$ in such a way that the
canonically normalized Boltzmann weights \eqref{renorm} have the
following quasi-classical asymptotics when
$\hbar\to0$,
\beq
\begin{array}{rcl}
\ds{\mathcal  W}(\alpha\,|\,\sigma_1,\sigma_2)&\!\!\!=\!\!\!&\ds\exp\Big\{
-\frac{{\mathcal L}(\alpha\,|\,\p_1,\p_2)}{\hbar} 
-{\mathcal L}_1(\alpha\,|\,\p_1,\p_2)+ O(\hbar)\Big\},
\\[.5cm]
\ds\sum_{\sigma}
\mathcal{S}(\sigma)\ldots&\!\!\!=\!\!\!&\ds\int \, \exp\Big\{-{\frac{{\mathcal
          C(\p)}}{\hbar}+{\mathcal C}_1(\p)}+O(\hbar)\Big\}\ldots\ \frac{d\p}
{\sqrt{2\pi \hbar}},
\end{array}\label{w-class}
\eeq
where $x_i$
denotes appropriately scaled
spin variables, $\p_i=\mbox{const }\,\sigma_i$, 
which in this limit, always become continuous.
The functions ${\mathcal L}(\alpha\,|\,\p_1,\p_2)$,
${\mathcal L}_1(\alpha\,|\,\p_1,\p_2) $, ${\mathcal C}(\p)$ and
${\mathcal C}_1(\p)$  are
independent of $\hbar$. As functions of complex variables, 
they are, in general, multi-valued functions
of the spins $x_1,x_2$ and the spectral parameter $\alpha$. 

One important thing which could be affected in the quasi-classical
limit is the assignment of the edge variables $\alpha_{(ij)}$ defined
by \eqref{difvar}. We
assume the rapidity variables $p_1,p_2,\ldots p_L$ remain unchanged
in the limit.  However the parameter $\cpar$ entering \eqref{difvar}
for the edges of the second type could depend on $\hbar$.  The models
discussed here fall into two different classes, where the parameter
$\cpar$ either
vanishes linearly in $\hbar$, or remains finite as $\hbar\to0$,
%
\beq \cpar_0=\lim_{\hbar\to0}\cpar=\left\{\begin{matrix}0,\qquad
&\mbox{(unphysical case),}\\[.2cm] >0,\qquad &\mbox{(physical case),}
\end{matrix}\right.\label{eta-clas}
\eeq
The discrete spin models (the Kashiwara-Miwa model and  its reductions)
fall into the first class, while all the continuous spin models
(including the master model, Faddeev-Volkov and Zamolodchikov's ``fishing-net'' models) belong to the
second class. As explained in Section~\ref{Zinvar} in the physical regime
({\it i.e.}, when all weights are positive) all the edge variables $\alpha_{(ij)}$
in \eqref{Z-def1} must be in the physical domain $0<\alpha_{(ij)}<\eta$.
However, if $\eta=O(\hbar)$, this domain shrinks to a point when
$\hbar\to0$. Therefore, unless all $\t_{ij}=0$ and the model is
trivial, there must be negative (or even complex) weights in the limit,
which is an unphysical regime from the point of statistical mechanics.

The symmetry of Boltzmann weights \eqref{sym} implies
\beq\label{sym-clas}
{\mathcal L}(\alpha\,|\,x,y)={\mathcal L}(\alpha\, |\,y,x)\;,
\eeq
while the inversion relations
\eqref{inv1} and \eqref{inv2} imply\footnote{Note that in
  \eqref{w-class} we use the canonically normalized Boltzmann weights
  \eqref{renorm} for which the factors $f(\pm\alpha)$ in \eqref{inv2}
  should be dropped. }
\beq
\begin{array}{rcl}
{\mathcal L}(\alpha\,|\,x,y)+{\mathcal L}(-\alpha\,|\,x,y)&\!\!\!=\!\!\!&0\,,\\[.3cm]
{\mathcal L}(\eta_0+\alpha\,|\,x,y)+{\mathcal
  L}(\eta_0-\alpha\,|\,x,y)&\!\!\!=\!\!\!&-{\mathcal C}(x)-{\mathcal C}(y)\,.
\end{array}\label{inv-clas}
\eeq
The above relations hold provided one chooses appropriate branches of
the function ${\mathcal L}(\alpha\,|\,x,y)$, if the argument 
$\alpha$ lies outside the physical domain $0<\alpha<\eta_0$.    
Note also, that for the unphysical regime, when $\eta_0=0$, the
function ${\mathcal  C}(x)$ must vanish, so that the two relations coincide. 

Substituting \eqref{w-class} into the partition function \eqref{Z-def1} one
obtains,
\beq
{\mathcal Z}=\int
\exp\Big\{\!-{\mathcal A}(\p)/\hbar-{\mathcal
  B}(\p)+O\,(\hbar)\Big\}\,
\prod_{i} \frac{d \p_i}{\sqrt{2\pi\hbar}}
\ ,\qquad \hbar\to0,\label{Zas-def}
\eeq
where the product is over all internal sites $i \in V_{int}({\mathscr
  G})$ of our graph $\mathscr{G}$, and
\bea
{\mathcal A}(\p)&\!\!\!=\!\!\!\!\!&\sum_{(ij)\in E(\mathscr G)}
{\mathcal L}(\alpha_{(ij)}\,|\,\p_i,\p_j)+\sum_{i\in
  V({\mathscr G})} {\mathcal C}(\p_i),\label{action}\\[.4cm]
{\mathcal B}(\p)&\!\!\!=\!\!\!\!\!&\sum_{(ij)\in E(\mathscr G)}
{\mathcal L}_1(\alpha_{(ij)}\,|\,\p_i,\p_j)
+\sum_{i\in V({\mathscr G})}{\mathcal C}_1(\p_i)+\mbox{terms proportional
  $\partial \cpar/\partial\hbar$},\label{action1}
\eea
and the sums are taken over all edges and over all sites of ${\mathscr
G}$.  As before, the external spins are kept fixed. The variables
$\alpha_{(ij)}$ in \eqref{action} and \eqref{action1} are defined by
\eqref{difvar} with $\cpar$ substituted by $\ocpar$, given in
\eqref{eta-clas}.
Calculating the integral \eqref{Zas-def} by the saddle point method one
obtains\footnote{%
Note, that the saddle point method requires some modifications
if the equations \eqref{motion} are defined to a continuous set of stationary
points. This situation occurs in the Zamolodchikov's ``fishing net''
model, which will be considered separately in
Section~\ref{sec:fishnet}.
}
\beq
\log{\mathcal Z}=-\frac{1}{\hbar}\,{\mathcal A}(\p^{(cl)})+
{\mathcal B}{(\p^{(cl)})}-\hf\log \det\Big\Vert\frac{\partial^2
  {\mathcal A}(\p)}{\partial
  \p_i\partial\p_j}\Big\Vert_{\p=\p^{(cl)}}+O(\hbar) \ .\label{Zexp}
\eeq
The symbol $\p^{(cl)}$ denotes the stationary point of the action
${\mathcal A}(\p)$, determined by the classical equations of motion
\beq
\frac{{\partial {\mathcal A}}(\p)}{\partial
  \p_i}\Big\vert_{\p=\p^{(cl)}}
=0,\qquad i\in
  V_{int}({\mathscr G}) \ .\label{motion}
\eeq
Introducing a new function 
\beq
\psi(\t\,|\,x,y)=\frac{\partial}{\partial x}\Lcal(\t\,|\,x,y)\,.
\label{psi-def}
\eeq
and using \eqref{action} one can write \eqref{motion} explicitly as discrete
Laplace-type equations, 
\beq
\sum_{(ij)\in
  star(i)} {\psi}(\alpha_{(ij)}\,|\,x_i,x_j)+
\frac{\partial}{\partial x_i} {\mathcal C}(x_i)=0\,,\qquad i\in
  V_{int}({\mathscr G})\;, \label{Laplace}
\eeq
where ${star}(i)$ denotes the set of edges meeting at the site $i$. Note, that
the parameters $\alpha_{(ij)}$ entering this equation obeys the sum
rule \eqref{sumrule1}.

\subsection{Classical star-triangle relation and invariance of the action}

Let us now recall the $Z$-invariance properties of the partition
function \eqref{Z-def1}. Obviously, these properties must hold for
{\em each term of the quasi-classical expansion \eqref{Zexp}}.
In particular, the classical action ${\mathcal A}(x^{(cl)})$,
evaluated on solutions of the classical equation of motion
\eqref{motion} (the leading term in the expansion
\eqref{Zexp}), remains invariant with respect to the star-triangle
moves\footnote{%
As explained in Section~\ref{tilings} the star-triangle moves are
equivalent to the ``flipping cube'' moves of the quad-graph $\mathscr{L}^*$
which is the dual to $\mathscr{L}$.}  of the rapidity graph
$\mathscr{L}$, shown in Figure \ref{startriangle},
\beq
{\mathcal A}(x^{(cl)}) \;-\; \mbox{invariant under the star-triangle moves
  of ${\mathscr L}$}\ .\label{A-inv}
\eeq
This result was obtained in \cite{BMS07a} and illustrated on the example of 
the Faddeev-Volkov model. The arguments of \cite{BMS07a} are 
rather general and apply  with no modifications to all
integrable edge-interaction models, which admit the
quasi-classical limit \eqref{w-class}.
Indeed, all mathematical relations required for the invariance
\eqref{A-inv} arise 
automatically from the quasi-classical expansion of the star-triangle
relation. To be more precise this expansion generates an infinite
number of non-trivial relations, one relation in each order of $\hbar$.
The statement \eqref{A-inv} only requires the first of these relations
arising in the leading order in $\hbar$.

Substituting \eqref{w-class} into \eqref{str-def} and taking
into account \eqref{rfactor}, one obtains
\begin{equation}\label{str-semi}
\int \frac{d\p_0}{\sqrt{2\pi\hbar}} \ \exp\left\{ -\frac{1}{\hbar}
\,\mathcal{A}_{{\bigstar}}(\p) - \mathcal{B}_{\bigstar}(\p)\right\}\;=\;
\exp\left\{-\frac{1}{\hbar}\, \mathcal{A}_{\triangle}(\p) -
\mathcal{B}_{\triangle}(\p)+O(\hbar)\right\}\;,
\end{equation}
where
\bea
\mathcal{A}_{\bigstar}(\p)&\!\!\!=\!\!\!&\Lcal(\ocpar-\alpha_1\,|\,\p_0,\p_1)+
\Lcal(\ocpar-\alpha_2\,|\,\p_0,\p_2)
+\Lcal(\ocpar-\alpha_3\,|\,\p_0,\p_3)+{\mathcal C}(\p_0)
\;,\label{A-star}\\[.3cm]
\mathcal{A}_{\triangle}(\p)&\!\!\!=\!\!\!&\Lcal(\alpha_1\,|\,\p_2,\p_3)+
\Lcal(\alpha_2\,|\,\p_3,\p_1)
+\Lcal(\alpha_3\,|\,\p_1,\p_2)\;.\label{A-tri}
\eea
and
\beq
\t_1+\t_2+\t_3=\ocpar.\label{csum}
\eeq
The symbol $x$ stands for the set $x=(x_0,x_1,x_2,x_3)$.
Expressions for $\mathcal{B}_\bigstar$ and $\mathcal{B}_\triangle$
are defined in a similar way; they are just specializations of
\eqref{action1} for the star and triangular graphs in Figure \ref{startriangle}.
Evaluating the integral \eqref{str-semi} by the saddle point method
one immediately obtains two non-trivial identities
valid for arbitrary values of $\p_1,\p_2,\p_3$. In the leading order in
$\hbar$ one gets
\beq
\mathcal{A}_\bigstar (\p_0^{(cl)},\p_1,\p_2,\p_3) =
\mathcal{A}_\triangle(\p_1,\p_2,\p_3), \label{cstr}
\eeq
where $\p_0^{(cl)}$ is the stationary point of the integral in
\eqref{str-semi}, {\it i.e.}, the value of $\p_0$, which solves the equation
\begin{equation}
\frac{\partial{\mathcal A}_\bigstar(x)}{\partial \p_0}\Big|_{x_0=x_0^{(cl)}}
=0\ .\label{statpoint}
\end{equation}
In the order $O(\hbar^0)$ one gets \cite{BMS07a}
\beq
\Big\{\frac{1}{2}\,\log\frac{\partial^2 \mathcal{A}_{\bigstar}(\p)}{\partial
  \p_0^2}
+\mathcal{B}_\bigstar(\p)\Big\}\Big\vert_{\p_0=\p_0^{(cl)}}
=\mathcal{B}_\triangle(\p)\;.
\eeq
The last relation will not be used in what follows. It is presented
here just to illustrate that the star-triangle relation has a consistent
expansion in powers of $\hbar$.

From now on we will omit the superfix ``(cl)'' for the solution of
\eqref{statpoint} and assume $x_0\equiv x_0^{(cl)}$.
Writing
\eqref{cstr} in full one obtains the {\em classical star-triangle relation} 
\beq\begin{array}{l}
\Lcal(\ocpar-\alpha_1\,|\,\p_0,\p_1)+
\Lcal(\ocpar-\alpha_2\,|\,\p_0,\p_2)
+\Lcal(\ocpar-\alpha_3\,|\,\p_0,\p_3)+{\mathcal C}(x_0)\\[.4cm]
\ \ \ \ \ \ \ \ 
\phantom{\qquad\qquad\Lcal(\ocpar-\alpha_1\,|\,\p_0,\p_1)}=\Lcal(\alpha_1\,|\,\p_2,\p_3)+
\Lcal(\alpha_2\,|\,\p_3,\p_1)
+\Lcal(\alpha_3\,|\,\p_1,\p_2)\;.
\end{array}\label{cstr2}
\eeq
where, as before, the arguments $\alpha_1,\alpha_2,\alpha_3$ obey the
relation \eqref{csum}. 
The stationary point $x_0$ is determined by the equation 
\begin{subequations}\label{stat03}
\beq
\psi(\ocpar-\alpha_1\,|\,\p_0,\p_1)+\psi(\ocpar-\alpha_2\,|\,\p_0,\p_2)
+\psi(\ocpar-\alpha_3\,|\,\p_0,\p_3)+
\frac{\partial}{\partial x_0} {\mathcal C}(x_0)=0,\label{stat0}
\eeq
with $\psi(\alpha\,|\,x, y)$ defined in \eqref{psi-def}.
It is convenient to regard the last equation 
as a constraint on the four
variables $x_0,x_1,x_2,x_3$, rather than an equation for $x_0$.
The classical star-triangle relation \eqref{cstr2} 
holds as long as this constraint is satisfied.
Note that it
can be re-written in three other equivalent forms.
To do this one needs to differentiate \eqref{cstr2} with respect to
$x_1$, $x_2$ or $x_3$. There is no need to take into account the dependence of 
$x_0$ on $x_1,x_2,x_3$, since the expression \eqref{cstr2} is
stationary with respect to $x_0$.  As a result one obtains 
\bea\label{stat4}
\ds
\psi(\t_2\,|\,x_1,x_3)
+\psi(\t_3\,|\,x_1,x_2)-\psi(\t_2+\t_3\,|\,x_1,x_0)&\!\!\!=\!\!\!&0,\label{stat1}\\[.3cm]
\ds
\psi(\t_1\,|\,x_2,x_3)
+\psi(\t_3\,|\,x_2,x_1)-\psi(\t_1+\t_3\,|\,x_2,x_0)&\!\!\!=\!\!\!&0,\label{stat2}\\[.3cm]
\ds
\psi(\t_1\,|\,x_3,x_2)
+\psi(\t_2\,|\,x_3,x_1)-\psi(\t_1+\t_2\,|\,x_3,x_0)&\!\!\!=\!\!\!&0 \label{stat3}\ .
\eea
\end{subequations}
Note that the function $\psi(\t\,|x,y)$ 
satisfies a pair of functional equations\footnote{As noted
before ${\mathcal C}(x)\equiv0$ when $\eta_0=0$.}
\beq\begin{array}{rcl}
\psi(\t\,|\,x,y)+\psi(-\t\,|\,x,y)&\!\!\!=\!\!\!&0, \\[.3cm]
\psi(\ocpar+\t\,|\,x,y)+\psi(\ocpar-\t\,|\,x,y)&\!\!\!=\!\!\!&\ds
-\frac{\partial}{\partial x} {\mathcal C}(x)\,,\label{psi-rel}
\end{array}\eeq
which simply follow from \eqref{inv1} and \eqref{inv2}.

\subsection{Consistency around a cube}

\begin{figure}[ht]
\begin{center}
\setlength{\unitlength}{0.30mm}
\begin{picture}(340,140)
\Thicklines
\put(20,20)
 {\begin{picture}(100,100)
 \path(10,15)(10,85)
 \path (15,90)(85,90)
 \path (90,85)(90,15)
 \path (85,10)(15,10)
 \put(10,10){\circle*{10}}\put(10,90){\circle{10}}
 \put(90,10){\circle{10}}\put(90,90){\circle*{10}}
 \put(-5,-5){$x_0$}\put(95,-5){$x_2$}\put(-5,100){$x_1$}\put(95,100){$x_{12}$}
 \put(45,45){$Q_{12}$}
\thinlines
\put(-35,45){$p_1$}
\path(-20,50)(-15,50)
\put(-5,50){\path (-10,-4)(0,0)\path(-10,4)(0,0)\path(-10,4)(-10,-4)}
\path(-5,50)(35,50)
\path(70,50)(120,50)
\path(50,-20)(50,-15)
\put(50,-5){\path (-4,-10)(0,0)\path(4,-10)(0,0)\path(4,-10)(-4,-10)}
\path(50,-5)(50,35)
\path(50,70)(50,120)
\put(58,-20){$p_2$}
 \end{picture}}
 \put(220,20)
 {\begin{picture}(100,100)
\path(10,15)(10,85)
 \path (15,90)(85,90)
 \path (90,85)(90,15)
 \path (85,10)(15,10)
 \put(10,10){\circle{10}}\put(10,90){\circle*{10}}
 \put(90,10){\circle*{10}}\put(90,90){\circle{10}}
 \put(-5,-5){$x_0$}\put(95,-5){$x_2$}\put(-5,100){$x_1$}\put(95,100){$x_{12}$}
 \put(45,45){$\overline{Q}_{12}$}
\thinlines
\put(-35,45){$p_1$}
\path(-20,50)(-15,50)
\put(-5,50){\path (-10,-4)(0,0)\path(-10,4)(0,0)\path(-10,4)(-10,-4)}
\path(-5,50)(35,50)
\path(70,50)(120,50)
\path(50,-20)(50,-15)
\put(50,-5){\path (-4,-10)(0,0)\path(4,-10)(0,0)\path(4,-10)(-4,-10)}
\path(50,-5)(50,35)
\path(50,70)(50,120)
\put(58,-20){$p_2$}
 \end{picture}}
\end{picture}
\end{center}
\caption{Graphical representation of the equations \eqref{QQ-def}
  corresponding to two different types of quadrilaterals.}
\label{fig-two-quads}
\end{figure}
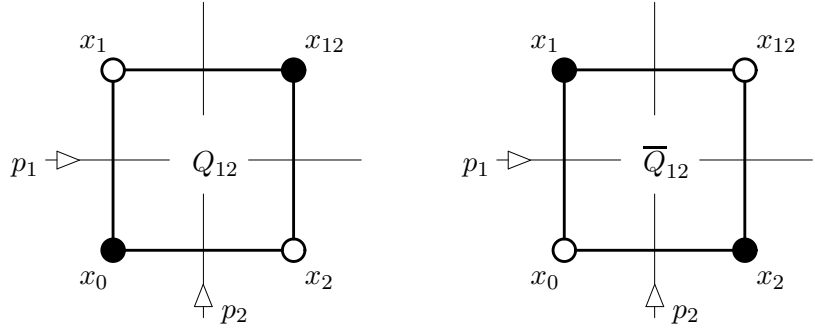

In \cite{AdlerBobenkoSuris} 
Adler, Bobenko and Suris introduced a remarkable class
of integrable discrete evolution equations.  
A distinguished feature of these
equations is that their integrability properties are automatically
satisfied due to the equations themselves (another way of describing
this situation would be to say that the corresponding ``Lax pair'' is
contained within the equations).  
The above equations are classical (not
quantum) evolution equations for a complex scalar field, defined on vertices
of a quad-graph. The later could be either a regular graph ({\it e.g.}, a
square lattice)  
or an irregular graph of type shown in Figure \ref{fig6}.
The four values of the field
$x_0,x_1,x_2,x_{12}$ at the vertices of an elementary quadrilateral,
as in Figure \ref{fig-two-quads},
are constrained by one relation $Q(x_0,x_1,x_2,x_{12})=0$. In general, 
this relation varies for different quadrilaterals (see below).    
The integrability conditions for
such system, are called the {\em consistency-around-a-cube}
conditions.  
The list of all solutions of these conditions for the case of
affine-linear constraints $Q(x_0,x_1,x_2,x_{12})$ admitting the symmetries of the square were found in
\cite{AdlerBobenkoSuris} (which we refer to below as the ABS list). 

Recently, Bobenko and Suris \cite{BS09}
have shown that every equation from that list corresponds to a
certain solution of the classical star-triangle relation. Previously
this fact was established \cite{BMS07a} for the Hirota equations, which is
$Q_{3,\delta=0}$ in the ABS list\footnote{%
This fact was also known to us
for $Q_{3,\delta=1}$ and $Q_4$ equations before the paper
\cite{BS09} has appeared.}. As shown in \cite{BS09}, a generic ABS
equation is related to a more general, than \eqref{cstr2}, 
classical star-triangle relation, containing different functions ${\mathcal
L}(\t\,|\,x_1,x_2)$ for different edges. In our setting this
corresponds to systems without the crossing symmetry.  

In \cite{BS09} solutions of the classical star-triangle relation were
obtained from solutions of the consistency-around-a-cube conditions.
Here we want to reverse the argument and consider a converse procedure. 
To do this we use an observation of \cite{BS09} that the constraints
\eqref{stat03}, associated with the classical star-triangle relation, 
can be identified with the so-called {\em three-leg
  form} \cite{AdlerBobenkoSuris} 
of the equation $Q(x_0,x_1,x_2,x_{12})=0$ on an
elementary quadrilateral.  
Note that the variable $x_0$ appears in every term of \eqref{stat0}.
For this reason we will 
call this equation the ``three-leg form centered at $x_0$". 
The other three equivalent forms of this relation \eqref{stat1}-\eqref{stat3}
are centered at $x_1$, $x_2$ and $x_3$, respectively.
 
The construction of a quad-graph ${\mathscr L}^*$ 
and its (oriented) rapidity graph ${\mathscr L}$
is explained in Sect.~\ref{tilings}. Here we assume the same notations. 
Recall that sites of ${\mathscr L}^*$ are colored 
black and white (every quad-graph is bipartite). There are two types
of quad-faces, differing by the position of white sites relative to
the directed rapidity lines as shown in Figure \ref{two-rhombi} and Figure \ref{fig-two-quads}. 
Let $x_0,x_1,x_2,x_{12}$ be the fields at the corners of a face,
and $p_1,p_2$ denote the rapidity variables arranged as
in  Figure \ref{fig-two-quads}. Define two different constraints $Q_{12}$ 
and $\overline{Q}_{12}$, 
\beq\begin{array}{l}
Q(p_1,p_2\,|\,x_0,x_1,x_2,x_{12})=\psi(p_1-p_2|\,x_2,x_1)+
\psi(p_2|\,x_2,x_0)-\psi(p_1|\,x_2,x_{12})\;,\\[.3cm]
\overline{Q}(p_1,p_2\,|\,x_0,x_1,x_2,x_{12})=
\psi(p_1-p_2|\,x_2,x_1)+\psi(\ocpar-p_1|\,x_2,x_{12})
-\psi(\ocpar-p_2\,|\,x_2,x_0), 
\end{array}\label{QQ-def}\eeq
where $\psi(\alpha\,|\,x,y)$ satisfies \eqref{stat03}. 
Below we will also use the abbreviated notations 
\beq
Q_{ij}(x_0,x_1,x_2,x_{12})\equiv Q(p_i,p_j\,|\,x_0,x_1,x_2,x_{12}),\qquad 
i,j=1,2,3
\eeq
and similarly for $\overline{Q}_{ij}$.

{\bf Consistency around a cube}. {\em Let the rapidity variables 
$p_1,p_2,p_3$ take arbitrary values and the fields 
$x,x',x_1,x_2,x_3,x_{12},x_{13},x_{23}$ be arranged as shown in
Figure \ref{fig-LHS}. Assume that the equations \eqref{stat03} are
satisfied. Then the system of three equations 
\begin{equation}\label{QQQ-lhs}
\overline{Q}_{12}(x,x_{23},x_{13},x_3)=0\;,\quad
Q_{13}(x_{12},x_2,x,x_{23})=0\;,\quad
\overline{Q}_{23}(x_1,x_{12},x_{13},x)=0\;,
\end{equation}
corresponding to the three faces of a cube, shown on the left side of 
Figure \ref{fig-LHS}, is consistent with the system of three equations 
\begin{equation}\label{QQQ-rhs}
Q_{12}(x_{12},x_{2},x_{1},x')=0\;,\quad
\overline{Q}_{13}(x_{1},x',x_{13},x_3)=0\;,\quad
Q_{23}(x',x_{2},x_{3},x_{23})=0\;.
\end{equation}
corresponding to the other three faces of a cube, shown on the right side of 
Figure \ref{fig-LHS}.}

\def\arrvert{{\path(0,-10)(0,0)\path(-5,0)(0,13)\path(5,0)(0,13)
\path(-5,0)(5,0)}}
\smallskip
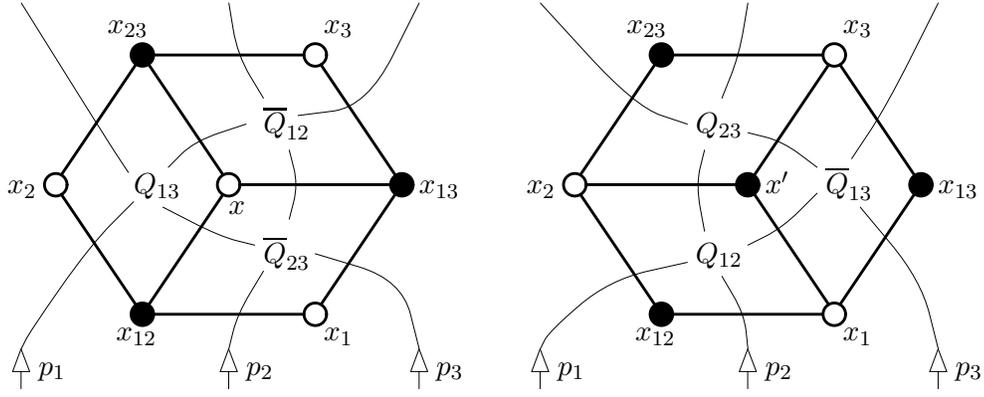
\begin{figure}[ht]
\begin{center}
\setlength{\unitlength}{0.23mm}
\begin{picture}(520,190)
\put(20,20)
 {\begin{picture}(200,150)
\Thicklines
 \path(3.25,80.63)(50,150)(143.5,150)
\path(153.25,144.37)(200,75)(153.25,5.63)
\path (143.5,0)(50,0)
\path(50,0)(3.25,69.37)
 \put(0,75){\circle{13}}\put(50,150){\circle*{13}}\put(150,150){\circle{13}}
 \put(200,75){\circle*{13}}\put(150,0){\circle{13}}\put(50,0){\circle*{13}}
 \put(155,-15){$x_{1}$}\put(-28,70){$x_{2}$}\put(35,-15){$x_{12}$}%
 \put(210,70){$x_{13}$}\put(155,163){$x_{3}$}\put(30,163){$x_{23}$}
 \put(100,75){\circle{13}}
\path(96.75,80.63) (50,150)
\path(106.5,75)(200,75)
\path(96.75,69.37)(50,0)
 \put(100,58){$x$}
 \put(45,70){$Q_{13}$}\put(120,105){$\overline{Q}_{12}$}
\put(120,30){$\overline{Q}_{23}$}%
\thinlines
\put(-20,-33)\arrvert
\spline(-20,-20)(-15,-10)(20,40)(45,65)
\spline (65,85)(80,100)(110,110)
\spline (140,115)(180,120)(210,180)
\put(100,-33)\arrvert
\spline(100,-20)(105,0)(120,25)
\spline(135,55)(140,75)(135,95)
\spline(120,125)(105,150)(100,180)
\put(210,-33)\arrvert
\spline(210,-20)(200,20)(150,35)
\spline(115,40)(95,45)(62,62)
\spline(42,82)(30,100)(-20,180)
\put(-10,-35){$p_1$}\put(110,-35){$p_2$}\put(220,-35){$p_3$}
 \end{picture}}
\put(320,20)
 {\begin{picture}(200,150)
 \Thicklines
 \path(3.25,80.63)(50,150)(143.5,150)
\path(153.25,144.37)(200,75)(153.25,5.63)
\path (143.5,0)(50,0)
\path(50,0)(3.25,69.37)
 \put(0,75){\circle{13}}\put(50,150){\circle*{13}}\put(150,150){\circle{13}}
 \put(200,75){\circle*{13}}\put(150,0){\circle{13}}\put(50,0){\circle*{13}}
 \put(155,-15){$x_{1}$}\put(-28,70){$x_{2}$}\put(35,-15){$x_{12}$}%
 \put(210,70){$x_{13}$}\put(155,163){$x_{3}$}\put(30,163){$x_{23}$}
 \put(100,75){\circle*{13}}
 \path(100,75)(146.75,144.37)
 \path(100,75)(146.75,5.63)
 \path(100,75)(6.5,75)
 \put(110,70){$x'$}
 \put(145,70){$\overline{Q}_{13}$}\put(70,105){$Q_{23}$}
\put(70,30){${Q}_{12}$} 
\thinlines

\put(-20,-33)\arrvert
\spline(-20,-20)(-15,-10)(20,20)(65,30)
\spline (100,40)(120,50)(140,70)
\spline (160,90)(180,120)(210,180)
\put(100,-33)\arrvert
\spline(100,-20)(98,0)(85,22)
\spline(75,48)(70,75)(75,95)
\spline(85,125)(95,150)(100,180)
\put(210,-33)\arrvert
\spline(210,-20)(200,20)(165,62)
\spline(140,85)(120,100)(100,105)
\spline(65,110)(35,120)(-20,180)
\put(-10,-35){$p_1$}\put(110,-35){$p_2$}\put(220,-35){$p_3$}
\end{picture}}
\end{picture}
\end{center}
\caption{Arrangement of the fields $x,x',x_1,x_2,x_3,x_{12},x_{13},x_{23}$
on the vertices of a cube.}
\label{fig-LHS}
\end{figure}
The proof is essentially identical to that of
\cite{AdlerBobenkoSuris}. Equations~\eqref{QQQ-lhs} contain three
relations for seven variables, leaving four degrees of freedom. For
example, if $x_1,x_{12},x_2,x_{23}$ are given, then $x,x_3,x_{13}$ are
uniquely determined. This fixes all variables entering
\eqref{QQQ-rhs}, except $x'$. To prove the consistency one needs to
show that each of the three relations in \eqref{QQQ-rhs} define the
same value of $x'$. For instance, suppose that $x'$ satisfies 
$\overline{Q}_{13}=0$. 
Combining this equation with two equations from \eqref{QQQ-lhs}, one obtains
\beq\begin{array}{c}
\overline{Q}_{13}(x_{1},x',x_{13},x_3)
-\overline{Q}_{12}(x,x_{23},x_{13},x_3)
-\overline{Q}_{23}(x_1,x_{12},x_{13},x)=\\[.3cm]
=Q(p_1-p_2,p_3-p_2\,|\,x_{12},x',x_{13},x_{23})=0
\end{array}
\eeq
Rewriting the last equation in the form centered at $x_{23}$, 
combining it with suitable forms of the equations $Q_{13}=0$ and
$\overline{Q}_{12}=0$ from \eqref{QQQ-lhs} (also centered at $x_{23}$)
and using the second relation in \eqref{psi-rel} one can readily deduce that 
the equation $Q_{23}=0$ in \eqref{QQQ-rhs} is satisfied. 
Similarly one can check that $Q_{12}=0$ as well, this time one
needs to use the first relation from \eqref{psi-rel}, with the relevant equations centered at $x_{12}$. 

{\bf Remark}. The above reasonings apply to two cases $\ocpar=0$ and
$\ocpar=\pi$. They correspond,  
respectively, to the unphysical and physical regimes from the point of  
the quasi-classical limit of a quantum model. 
Note that for $\ocpar=0$ the two constraints in \eqref{QQ-def}
coincide, thanks to \eqref{psi-rel}. Thus the
consistency equations \eqref{QQQ-lhs} and \eqref{QQQ-rhs} in this case
involve the only one constraint with different rapidity variables
for different quadrilaterals. 


\nsection{Particular lattice models and their quasiclassical limits}
\label{sec:models}

The models considered here (except for the gamma function model 
of Section \ref{sec:gammamodels}), possess all of the properties discussed in the
previous Section, including the rapidity difference property, crossing
symmetry, positivity, inversion relations \eqref{inv1}, \eqref{inv2}
and reflection symmetry \eqref{sym}.

\subsection{Master solution to the star-triangle relation}

The Boltzmann weights of the master solution are $\pi$-periodic, so spins take arbitrary values modulo $\pi$. It is convenient to regard them as
\begin{equation}
0\leq x_i < \pi\;.
\end{equation}
Let $q$ and $p$ be elliptic nomes,
\begin{equation}
q=e^{\ii\pi\tau}\,,\quad p=e^{\ii\pi\tau'}\;.
\end{equation}
The crossing parameter is defined by
\begin{equation}
\eta = \frac{\pi}{2\ii}(\tau+\tau')\;.
\end{equation}
In what follows, we use the standard notations \cite{WW} for Jacobi $\vartheta$-functions, {\it e.g.}
\begin{equation}
\vartheta_1(z\,|\,\tau) = 2q^{1/4}\sin(z)\prod_{n=1}^\infty (1-e^{2\ii z} q^{2n})(1-e^{-2\ii z}q^{2n})(1-q^{2n})\;.
\end{equation}
Explicit expressions for the Boltzmann weights contain two special functions. The elliptic $\Gamma$-function is defined by
\begin{equation}\label{egamma}
\Gamma(z) = \prod_{n,m=0}^\infty \frac{1-z^{-1} q^{2n+2} p^{2m+2}}{1-zq^{2n}p^{2m}}\;,
\end{equation}
however, a more convenient notation is
\begin{equation}\label{Phi}
\Phi(z)=\Gamma(e^{-2\ii(z-\ii\eta)}) = \exp\left\{\sum_{n\neq 0} \frac{e^{-2\ii z n}}{n(q^n-q^{-n})(p^n-p^{-n})}\right\}\;,
\end{equation}
since $\Phi(z)\Phi(-z)=1$. Another special function is
\begin{equation}\label{Kappa}
K_e(\alpha)=\exp\left\{\sum_{n\neq 0}\frac{e^{4\alpha n}}{n(q^n-q^{-n})(p^n-p^{-n})(q^np^n+q^{-n}p^{-n})}\right\}\;.
\end{equation}
The function $K_e(\alpha)$ satisfies the functional relations
\begin{equation}
\frac{K_e(\eta-\alpha)}{K_e(\alpha)}=\Gamma(e^{-4\alpha})\;,\quad K_e(\alpha) K_e(-\alpha)=1\;.
\end{equation}
These equations correspond to (\ref{func2}) with $f(\alpha)=\Gamma(e^{-4\alpha})$ and $\kappa_s=1$.
The canonically normalized Boltzmann weight is defined by
\begin{equation}\label{q4wts}
\mathcal{W}(\alpha\,|\,x,y)=
K_e(\alpha)^{-1} \frac{\Phi(x-y+\ii\alpha)}{\Phi(x-y-\ii\alpha)}
\frac{\Phi(x+y+\ii\alpha)}{\Phi(x+y-\ii\alpha)}\;.
\end{equation}
This Boltzmann weight admits the symmetries:
\begin{equation}\label{q4sym}
\mathcal{W}(\alpha\,|\,x,y)=\mathcal{W}(\alpha\,|\,y,x)=\mathcal{W}(\alpha\,|\,\pm x,\pm y)\;.
\end{equation}
The one-point weight is defined by
\begin{equation}
\mathcal{S}(x)=\frac{1}{2\pi}\, e^{\eta/2} \vartheta_1(2x\,|\,\tau)\,\vartheta_1(2x\,|\,\tau')\;.
\end{equation}
The weights are $\pi$-periodic with respect to $x,y$. The weights are positive when $\eta$ is real and $0\leq\alpha\leq\eta$.
The weights satisfy the difference relations
\begin{equation}
\frac{\mathcal{W}(\alpha\,|\,x-\frac{\pi\tau'}{2},y)}{\mathcal{W}(\alpha\,|\,x+\frac{\pi\tau'}{2},y)} = 
\frac{\vartheta_4(x-y+\ii\alpha\,|\,\tau)}{\vartheta_4(x-y-\ii\alpha\,|\,\tau)}
\frac{\vartheta_4(x+y+\ii\alpha\,|\,\tau)}{\vartheta_4(x+y-\ii\alpha\,|\,\tau)}\;,\quad \textrm{and similar with} \;\;
\tau \leftrightarrow \tau'\;.
\end{equation}
The weights have the canonical normalisation
\begin{equation}
\mathcal{W}(0\,|\,x,y)=1\;,\quad \mathcal{W}(\eta\,|\,x,y)=\frac{1}{2\mathcal{S}(x)}\left(\delta(x-y)+\delta(x+y)\right)\;,
\end{equation}
and satisfy the corresponding inversion relations
\begin{equation}
\mathcal{W}(\alpha\,|\,x,y)\,\mathcal{W}(-\alpha\,|\,x,y)=1\;,
\end{equation}
and
\begin{equation}
\int_0^\pi dz\, \mathcal{S}(z)\,\mathcal{W}(\eta-\alpha\,|\,x,z)\,\mathcal{W}(\eta+\alpha\,|\,z,y) = \frac{1}{2\mathcal{S}(x)}\,(\delta(x-y)+\delta(x+y))\;.
\end{equation}
Finally the Boltzmann weights satisfy the star-triangle equation
\begin{equation}\label{masterste}
\begin{array}{r}
\ds\int_{-\pi/2}^{\pi/2} dx\,\mathcal{S}(x)\,
\mathcal{W}(\eta-\alpha_1\,|\,x_1,x)\,\mathcal{W}(\alpha_1+\alpha_3\,|\,x_2,x)\,\mathcal{W}(\eta-\alpha_3\,|\,x_3,x) \\[0.3cm]
\ds=
\mathcal{W}(\alpha_1\,|\,x_2,x_3)\,
\mathcal{W}(\eta-\alpha_1-\alpha_3\,|\,x_1,x_3)\,
\mathcal{W}(\alpha_3\,|\,x_1,x_2)\;.
\end{array}
\end{equation}

\subsubsection{Classical limit of the master solution}

The classical limit is the limit $p\to 1$. Let $\tau'=\ii \hbar/\pi$, $\hbar\to 0$. In this limit
\begin{equation}
\Phi(z) = \exp\left\{ -\frac{1}{\hbar} \lambda_4(z\,|\,\tau) + O(1)\right\}\;,
\end{equation}
and
\begin{equation}
K(\alpha)=\exp\left\{-\frac{1}{\hbar} \lambda_4(2\ii\alpha\,|\,2\tau) + O(1)\right\}\;,
\end{equation}
where
\begin{equation}
\lambda_4(z\,|\,\tau)=\frac{1}{\ii} \int_0^z dx \log \overline{\vartheta}_4(x\,|\,\tau)\;,\quad 
\overline{\vartheta}_4(x\,|\,\tau)=\prod_{n=1}^\infty (1-\EXP^{2\ii x} q^{2n+1})(1-\EXP^{-2\ii x} q^{2n+1})\;.
\end{equation}
$\lambda_4(z\,|\,\tau)$ is even and $\pi$-periodic. The two-point Lagrangian
\begin{equation}
\mathcal{W}(\alpha\,|\,x,y) = \exp\left\{-\frac{1}{\hbar} \mathcal{L}(\alpha\,|\,x,y) + O(1)\right\}\;,
\end{equation}
and one-point Lagrangian
\begin{equation}
\mathcal{S}(x) = \exp\left\{-\frac{1}{\hbar}\, \mathcal{C}(x) + O(\log \hbar)\right\}\;,
\end{equation}
are given by
\begin{equation}
\mathcal{L}(\alpha\,|\,x,y) = \lambda_4(x-y+\ii\alpha)-\lambda_4(x-y-\ii\alpha) + \lambda_4(x+y+\ii\alpha) -\lambda_4(x+y-\ii\alpha) -\lambda_4(2\ii\alpha\,|\,2\tau)\;,
\end{equation}
and
\begin{equation}
\mathcal{C}(x) = (2|x|-\frac{\pi}{2})^2\;,\quad |x|<\frac{\pi}{2}\;,
\end{equation}
respectively. An equivalent expression for $\mathcal{L}_\alpha(x,y)$ is
\begin{equation}
\mathcal{L}(\alpha\,|\,x,y)=\frac{1}{\ii}\int_0^{x-y}dz\log\frac{\vartheta_4(z+\ii\alpha\,|\,\tau)}{\vartheta_4(z-\ii\alpha\,|\,\tau)} + \frac{1}{\ii}\int_{\pi/2}^{x+y}dz\log\frac{\vartheta_4(z+\ii\alpha\,|\,\tau)}{\vartheta_4(z-\ii\alpha\,|\,\tau)}\;.
\end{equation}
The crossing parameter in the classical limit is
\begin{equation}
\eta_0=\frac{\pi\tau}{2\ii}\;.
\end{equation}
Thus the classical limit corresponds corresponds to the equations labeled as $Q_4$ in  \cite{AdlerBobenkoSuris} with $\eta_0\neq 0$,

\subsection{Kashiwara-Miwa model}\label{sec:kmmodel}

In 1986 Kashiwara and Miwa \cite{Kashiwara:1986} found an elliptic
solution of the star-triangle relation ({\it i.e.}, parameterized by elliptic
functions of the rapidity variable) where spins take $N\ge2$ distinct
values
\beq
a,b,c,\ldots\in\{0,1,\ldots N-1\}, \qquad N\ge2.
\eeq
Their model contains the Ising model as the special case $N=2$. The Kashiwara-Miwa model can also be derived from the master solution in the limit
\begin{equation}
p\to \EXP^{\ii\pi/N}\;,
\end{equation}
for details see the reference \cite{Bazhanov:2010kz}.  The model contains three parameters:
\beq
N\ge2,\qquad \zeta\in {\mathbb Z},\qquad \tau\in {\mathbb C}, \quad
\im \tau>0 .
\eeq
Define the two functions
\beq
\ds r(\alpha\,|\,n)=  \ds\prod_{j=1}^{n} \,
\frac{\vt_1 \big(\cpar\,(j-\hf)-\hf\,\alpha\,\big|\,\tau\,\big)_{\phantom{|}}}
{\vt_1
  \big(\cpar\,(j-\hf)+\hf\,\alpha\,\big|\,\tau\,\big)^{\phantom{|}}}\ ,\qquad
\ds t(\alpha\,|\,n) =  \ds\prod_{j=1}^{n} \,
\frac{\vt_4 \big(\cpar\,(j-\hf)-\hf\,\alpha\,\big|\,\tau\,\big)_{\phantom{|}}}
{\vt_4
  \big(\cpar\,(j-\hf)+\hf\,\alpha\,\big|\,\tau\,\big)^{\phantom{|}}} \;,
\label{rt-def}
\eeq
where
\beq
\cpar=\pi/{N},  \label{eta-def}
\eeq
is the crossing parameter which enters \eqref{difvar}.
The weights of the Kashiwara-Miwa model read
\beq
W(\alpha\,|\,a,b)=
r(\alpha\,|\,a-b)\, t(\alpha\,|\,a+b+\zeta),\quad
S(a)=\frac{\vt_4(\cpar(2a+\zeta)\,|\,\tau)}{\vt_4(0\,|\,\tau)}\ ,\label{KM-def}
\eeq
where $a,b$ are integer spins.
Note that the functions \eqref{rt-def}
are periodic in their second argument
\beq
r(\alpha\,|\,n)= r(\alpha\,|\,n+N),\qquad
t(\alpha\,|\,n)= t(\alpha\,|\,n+N)\ ,
\eeq
therefore the weights \eqref{KM-def}
are periodic with respect to shifts of spins,
\beq
W(\alpha\,|\,a,b)=W(\alpha\,|\,a+N,b)=W(\alpha\,|\,a,b+N),\qquad S(a+N)=S(a)\ .
\label{spin-per}
\eeq
Next, the function $r(\alpha\,|\,n)=r(\alpha\,|-n)$ is
an even function of $n$, therefore the weights are unchanged by interchanging
the spins $a$ and $b$,
\beq
W(\alpha\,|\,a,b)=W(\alpha\,|\,b,a)\ .\label{sym2}
\eeq
The weights are real and
positive when $\re \tau=0$ and $\alpha$ is real and lies in the
interval $0 < \alpha< \cpar$.

An explicit expression for the factor
$f(\alpha)$ in this case was conjectured in \cite{Bax02rip}
\beq
f(\alpha)
 =  \ds \zsite\,\prod_{j=1}^{[N/2]} \,
\frac{\vt_1 \big(\cpar\,(j-\hf)+\hf\,\alpha\,\big|\,\hf\tau\big)_{\phantom{|}}}
{\vt_1 \big(\cpar\,j-\hf\,\alpha\,\big|\,\hf\tau\big)^{\phantom{|}}},
\qquad \zsite=g\,\sqrt{N},
\eeq
where $[N/2]$ denotes the integer part of $N/2$ and the constant
\beq
g=\prod_{m=1}^\infty \left(\frac{1+q^m}{1-q^m}\right)
\left(\frac{1-q^{Nm}}{1+q^{Nm}}\right),\qquad  q=e^{i\pi\tau}\;,
\eeq
is determined from the requirement $f(0)=1$.

Solving \eqref{func2} under the
assumptions that $\re \tau=0$ and that
$\log \kappa_e(\alpha)$ analytic and bounded in the
rectangle $0\le \re \alpha \le \cpar$, \ $0\le \im \alpha \le \im \tau$,
one obtains
\beq
\log \kappa_e(\alpha)\;=\; -\ii\, \frac{N-1}{4\tau}\,\alpha -
\sum_{k=1}^\infty \; \frac{1}{k} \;
\frac{\tilde{q}^{k}-\tilde{q}^{Nk}}{(1+\tilde{q}^{k})\,(1+\tilde{q}^{Nk})} \;
\frac{w^{2k}-w^{-2k}}{\tilde{q}^{k}-\tilde{q}^{-k}}\;,\label{z-km}
\end{equation}
where $w$ and $\tilde{q}$ are defined by
\beq
 w=\exp(-i{\alpha}/\tau),\qquad \tilde{q}=\exp(-2i{\cpar}/\tau),
\qquad |\tilde{q}|<1\ .\label{wq-def}
\eeq
The expression \eqref{z-km} can be written in the exponential form
\begin{equation}
\kappa_e(\alpha)\;=\;\EXP^{-\ii(N-1)\alpha/4\tau}\;{\mathsf{z}_1(w)}
/{\mathsf{z}_N(w)}\;,\label{z-exp}
\end{equation}
where
\begin{equation}
\mathsf{z}_N(w)\;=\;\prod_{j=0}^{N-1}\prod_{k=1}^\infty \left[
\frac{(1-\tilde{q}^{2kN+2j+1}w^{+2})\,(1-\tilde{q}^{2kN+2j-N+1}w^{-2})}
{(1-\tilde{q}^{2kN+2j+1}w^{-2})\,(1-\tilde{q}^{2kN+2j-N+1}w^{+2})}
\right]^k\;,\label{zn-def}
\end{equation}
and ${\mathsf z}_1(w)$ is defined by the last formula with $N=1$.
Note, in particular, that for the Ising model, $N=2$, one obtains
\beq
\kappa_e(\alpha)=
\EXP^{-\ii\alpha/4\tau}\;{\mathsf{z}_1(w)}
/{\mathsf{z}_2(w)}\;,\qquad N=2, \label{z-ising}
\eeq
where
\begin{equation}
\mathsf{z}_1(w)=\prod_{n=1}^\infty
\left[\frac{(1-\tilde{q}^{2n}w^{-2})(1-\tilde{q}^{2n+1}w^{+2})}{(1-\tilde{q}^{2n}
w^{+2})(1-\tilde{q}^{2n+1}w^{-2})}\right]^n,\qquad {\mathsf z}_2(w)=
\prod_{n=1}^\infty
\frac{(1-\tilde{q}^{4n-1}w^{-2})}{(1-\tilde{q}^{4n-1}w^{+2})} \ .\label{z12-def}
\end{equation}

\subsubsection{Classical limit of the Kashiwara-Miwa model}

The classical limit for the Kashiwara-Miwa model corresponds to $N\to\infty$. Let
\beq \hbar = \frac{2\pi}{N}\;,\quad x=\hbar \sigma\;, \eeq
where $\sigma$ represents the original discrete spin, and $x$ is the spin in the continuous limit.
In the limit $\hbar=2\eta\to 0$ the condition $0<\alpha<\eta$ makes the explicit classical limit trivial. Moreover, the regime of real $\alpha$ becomes ill-defined since the poles of the Boltzmann weight condense to a branch along the real axis. However, the case of imaginary spectral parameter is well defined. Changing then $\alpha\to\ii\alpha$, one obtains in the limit $\hbar\to 0$
\begin{equation}\label{w-i-class}
\ds \mathcal{W}(\ii\alpha\,|\,\frac{x}{\hbar},\frac{y}{\hbar})\; \sim\; \exp\left\{ -\frac{\ii}{\hbar} \mathcal{L}(\ii\alpha\,|\,x,y)\right\}\;,\quad
%
%
\overline{\mathcal{W}}(\ii\alpha\,|\,\frac{x}{\hbar},\frac{y}{\hbar})\; \sim\; \exp\left\{ -\frac{\ii}{\hbar} \overline{\mathcal{L}}(\ii\alpha\,|\,x,y)\right\}\;,
\end{equation}
where
\begin{equation}\label{overL}
\overline{\mathcal{L}}(\ii\alpha\,|\,x,y)\;=\;\mathcal{L}(-\ii\alpha\,|\,x,y)\;=\;-\mathcal{L}(\ii\alpha\,|\,x,y)\;,
\end{equation}
and $\mathcal{L}(\ii\alpha\,|\,x,y)=\mathcal{L}(\ii\alpha\,|\,y,x)$ is given by
\begin{equation}
\begin{array}{l}
\ds \mathcal{L}(\ii\alpha\,|\,x,y) \;=\; \ii \int_0^{x-y}\;
\log\frac{\vartheta_1(\hf (z-\ii\alpha)\,|\,\tau)}{\vartheta_1(\hf
(z+\ii\alpha)\,|\,\tau)}\;dz\;+\;\ii\int_0^{x+y}\; \log\frac{\vartheta_1(\hf
(z-\ii\alpha)\,|\,\tau)}{\vartheta_1(\hf
(z+\ii\alpha)\,|\,\tau)}\;dz\\
[7mm]
\ds\phantom{xxxxxxxxxx}
+\;\int_0^\alpha \log\left|\frac{\ds\vartheta_1(\ii\frac{z}{2}\,|\,\frac{\tau}{2})}{\ds\vartheta_2(\ii\frac{z}{2}\,|\,\frac{\tau}{2})}\right| \;dz\;.
\end{array}
\end{equation}
The first integral here must be understood as
\begin{equation}
\ii\int_0^x \log\frac{\vartheta_1(\hf (z-\ii\alpha)\,|\,\tau)}{\vartheta_1(\hf
(z+\ii\alpha)\,|\,\tau)}\;dz\; = \pi|x| - \frac{x^2}{2} + \ii \int_0^x \log\frac{(\EXP^{-\alpha-\ii z};q^2)_\infty (\EXP^{\alpha+\ii z}q^2;q^2)_\infty}{(\EXP^{-\alpha+\ii z};q^2)_\infty (\EXP^{\alpha-\ii z}q^2;q^2)_\infty}\, dz\;,
\end{equation}
where 
\begin{equation}
|x|\;\leq\; 2\pi\;,\quad 0<\alpha\;.
\end{equation}
The one-point Lagrangian is zero, $\mathcal{C}(x)=0$. The Lagrangians are canonically normalized and correspond to $Q_4$ with $\eta_0=0$.

Note that the condition (\ref{overL}) is assumed for all classical models with $\eta_0=0$.

\subsection{Hyperbolic limit of the master solution}

The hyperbolic limit of the master solution, is the limit when $q,p\to 1$,
\begin{equation}
\tau=\ii\,\frac{b}{T}\;,\quad \tau'=\ii\,\frac{b^{-1}}{T}\;,\quad T\to \infty\;.
\end{equation}
Such a limit was first considered by Spiridonov \cite{Spiridonov3} and details of this limit can be found in Appendix. 
There are two regimes of spin variables providing two hyperbolic solutions of the following star-triangle equation
\begin{equation}\label{str-hyperbolic}
\begin{array}{r}
\ds\int_{-\infty}^\infty dx \,\mathcal{S}(x)\,
\mathcal{W}(\eta-\alpha_1\,|\,x_1,x)\,\mathcal{W}(\alpha_1+\alpha_3\,|\,x_2,x)\,\mathcal{W}(\eta-\alpha_3\,|\,x_3,x) \\[0.3cm]
\ds =
\mathcal{W}(\alpha_1\,|\,x_2,x_3)\,
\mathcal{W}(\eta-\alpha_1-\alpha_3\,|\,x_1,x_3)\,
\mathcal{W}(\alpha_3\,|\,x_1,x_2)\;.
\end{array}
\end{equation}
In this section set the crossing parameter to be
\begin{equation}
\eta = \frac{1}{2}(b+b^{-1})\;,\quad \eta > 0\;.
\end{equation}
The spins now take values $x_i\in\mathbb{R}$, and spectral parameters are restricted to $0<\alpha_i<\eta$.  It is convenient to use symmetric dilogarithm function
\begin{equation}\label{fi-def}
\phi(z) = \exp\left\{\frac{1}{4} \int_{pv} \frac{\EXP^{-2\ii z w}}{\sinh(bw)\sinh(b^{-1}w)}\frac{dw}{w}\right\}\;,
\end{equation}
which satisfies $\phi(z)\phi(-z)=1$, and symmetric normalization function
\begin{equation}\label{fvkappa}
\kappa_e(\alpha)=\exp\left\{\frac{1}{8} \int_{pv} \frac{\EXP^{4\alpha w}}{\sinh(bw)\sinh(b^{-1}w)\cosh((b+b^{-1})w)} \frac{dw}{w}\right\}\;,
\end{equation}
where $\int_{pv}$ denotes the principal value integral.  The function $\kappa_e(\alpha)$ is a solution of (\ref{func2}) with $f(\alpha)=\phi(\ii\eta-2\ii\alpha)$:
\begin{equation}
\frac{\kappa_e(\eta-\alpha)}{\kappa_e(\alpha)}=\phi(\ii\eta-2\ii\alpha)\;,\quad \kappa(\alpha)\kappa(-\alpha)=1\;.
\end{equation}

The first hyperbolic solution to the star-triangle equation is given by
\begin{equation}\label{Q3delta1}
\mathcal{W}(\alpha\,|\,x,y) = \frac{1}{\kappa_e(\alpha)} \frac{\phi(x-y+\ii\alpha)}{\phi(x-y-\ii\alpha)} \frac{\phi(x+y+\ii\alpha)}{\phi(x+y-\ii\alpha)}\;,
\end{equation}
with
\begin{equation}
\mathcal{S}(x)=2\sinh(2\pi b x)\sinh(2\pi b^{-1} x)\;.
\end{equation}
The Boltzmann weights \eqref{Q3delta1} retain the symmetries \eqref{q4sym}.  This solution corresponds to the regime of small external spins in (\ref{masterste}),
\begin{equation}
x_i\to \frac{\pi}{T}\,x_i\;.
\end{equation}
In this case only when the integrand is near zero
\begin{equation}
x\to \frac{\pi}{T}\, x\;,
\end{equation}
will there be a contribution to the integral in the left hand side of (\ref{masterste}).

\subsubsection{Faddeev-Volkov model}
Another solution of the hyperbolic star-triangle equation corresponds to the Boltzmann weights for the Faddeev-Volkov model. In 1995 Faddeev and Volkov obtained \cite{FV95} a solution of the star
triangle relation which, in some sense, could be regarded as an
analytic continuation the Fateev-Zamolodchikov solution to negative
number of spin states $N$. Remarkably, the correspoding model of
statistical mechanics has positive Boltzmann weights\cite{BMS07a}, its
partition function in the large-lattice limit was calculated in
\cite{BMS07a,BMS07b}. The Boltzmann weights for the Faddeev-Volkov model are given by
\begin{equation}\label{FV}
\mathcal{W}(\alpha\,|\,x,y) = \frac{1}{\kappa_e(\alpha)} \frac{\phi(x-y+\ii\alpha)}{\phi(x-y-\ii\alpha)}\;,\quad 
\mathcal{S}(x)=1\;,
\end{equation}
where $\kappa_e(\alpha)$ is defined by (\ref{fvkappa}). It corresponds to the regime of external spins in (\ref{masterste})
\begin{equation}
x_i\to \frac{\pi}{4} + \frac{\pi}{T}x_i\;,
\end{equation}
so that only vicinities of $x\sim \pm \pi/4$,
\begin{equation}
x\to \pm\frac{\pi}{4} + \frac{\pi}{T}x\;,
\end{equation}
contribute to the integral.

The Boltzmann weights of the Faddeev-Volkov model are symmetric, 
\begin{equation}
\mathcal{W}(\alpha\,|\,x,y)\; =\; \mathcal{W}(\alpha\,|\,y,x)\;,
\end{equation} 
and possess a {\it self-duality property},
\begin{equation}
\mathcal{W}(\eta-\alpha\,|\,x,y) = \int_{\mathbb{R}} e^{2\pi\ii(x-y)z}\, \mathcal{W}(\alpha\,|\,z,0)\;.
\end{equation}

\subsubsection{Classical limit of the hyperbolic models}

The classical limit implies the re-scale of the variables
\begin{equation}
\alpha\to \frac{\alpha}{\pi b}\;,\quad x\to \frac{x}{\pi b}\;,\quad \hbar = \pi b^2\;.
\end{equation}
with $b\rightarrow0$.  The two-point Lagrangian for (\ref{Q3delta1}) is
\begin{equation}
\mathcal{L}(\alpha\,|\,x,y) = \frac{1}{\ii} \left(\int_0^{x-y} \log\frac{\cosh(z+\ii\alpha)}{\cosh(z-\ii\alpha)}dz +
\int_0^{x+y} \log\frac{\cosh(z+\ii\alpha)}{\cosh(z-\ii\alpha)}dz\right) + 2\int_0^{\alpha}\log2\cos(z)dz\;,
\end{equation}
and the one-point Lagrangian is
\begin{equation}
\mathcal{C}(x)=-2\pi|x|\;.
\end{equation}
The crossing parameter in the classical limit is $\eta_0=\pi/2$.

The two-point Lagrangian for the Faddeev-Volkov model (\ref{FV}) in the classical limit is
\begin{equation}
\mathcal{L}(\alpha\,|\,x,y)=\frac{1}{\ii} \int_0^{x-y} \log\frac{\cosh(z+\ii\alpha)}{\cosh(z-\ii\alpha)} dz +2\int_0^\alpha \log 2\cos(z)dz \;.
\end{equation}
The one-point Lagrangian is zero.

Both hyperbolic models \eqref{Q3delta1}, \eqref{FV}, in the classical limit correspond to $Q_{3,\delta=1}$ and $Q_{3,\delta=0}$ with $\eta_0=\pi/2$ respectively.

\subsection{Fateev-Zamolodchikov $Z_N$-model}

Taking a straightforward trigonometric limit, $\im \tau\to +\infty$, in the
Kashiwara-Miwa model one obtains from \eqref{KM-def}
\beq
\ds W(\alpha\,|\,a,b) =  \ds\prod_{j=1}^{a-b} \,
\frac{\sin \big(\cpar\,(j-\hf)-\hf\,\alpha\,\big)_{\phantom{|}}}
{\sin \big(\cpar\,(j-\hf)+\hf\,\alpha\,\big)^{\phantom{|}}},\qquad
 S(a)\equiv 1\;,\quad \eta=\frac{\pi}{N} \ .\label{FZ-def}
\eeq 
The model contains only one integer parameter $N\ge2$. As
before the spins take the values $a,b=0,1,\ldots,N-1$ and the
crossing parameter, $\cpar=\pi/N$, takes the same value as in
\eqref{eta-def}. The
resulting model is exactly the Fateev-Zamolodchikov model
\cite{FZ82} obtained in 1982. Obviously, the weights
\eqref{FZ-def} retain the symmetries \eqref{spin-per} and
\eqref{sym2}. They also acquire an additional $Z_N$-symmetry as
they only depend on the difference of spins $a-b\pmod N$.
The factor \eqref{fpq-def} for this case was calculated in
\cite{Bax02rip},
\beq
f(\alpha)
 =  \ds \zsite\,\prod_{j=1}^{[N/2]} \,
\frac{\sin \big(\cpar\,(j-\hf)+\hf\,\alpha\,\big)_{\phantom{|}}}
{\sin \big(\cpar\,j-\hf\,\alpha\,\big)^{\phantom{|}}},\qquad
\zsite=\sqrt{N}\ . \eeq The weights are real and positive when $0
< \re \alpha< \cpar$.
Taking the limit of \eqref{z-km} when $\im \tau\to+\infty$, one obtains
\begin{equation}
\log\kappa_e(\alpha) = -\frac{1}{2}\int_{0}^{\infty}\;
\frac{\sinh(N-1)\pi x}{\cosh \pi x \cosh N \pi x}\;
\frac{\sinh 2N\alpha x}{\sinh 2\pi x}\; \frac{dx}{x} \ .
\end{equation}
Finally, note that the weights \eqref{FZ-def} are self-dual \cite{Bax82},
\beq
W(\cpar-\alpha\,|\,a,b)=\ds N^{-1}\,f(\t)\,\sum_{k=0}^{N-1}
\omega^{k\,(a-b)}\ W(\alpha\,|\,k,0), \qquad \omega=\EXP^{2\pi i/N}\
  .\label{fz-dual}
\eeq
The scalar factor in front of the sum can be calculated in the same
way as the factor ${\mathcal R}_{123}$ in
Section~\ref{factor-R}. Consider both sides of \eqref{fz-dual} as an
element $(a,b)$ of some matrix. Taking the determinant of this matrix
one immediately obtains \eqref{fpq-def}.

\subsubsection{Classical limit of the Fateev-Zamolodchikov model}

The classical limit can be taken similarly to the classical limit for Kashiwara-Miwa model (in fact, it is just the trigonometric limit). The regime of imaginary $\alpha$ and convention (\ref{w-i-class}) give the canonically normalized two-point Lagrangian
\begin{equation}
{\mathcal L}(\ii\alpha\,|\,x,y)\;=\;\ii\int_0^{x-y} \;
\log\frac{\sin(\hf (z-\ii\alpha))}{\sin(\hf (z+\ii\alpha))}dz \;+\;\int_0^\alpha \log\tanh\frac{z}{2}\,dz\;,
\end{equation}
where
\begin{equation}
\ii\int_0^x \log\frac{\sin(\hf (z-\ii\alpha))}{\sin(\hf (z+\ii\alpha))}dz \;=\; \pi|x|-\frac{x^2}{2} - 2\int_0^x \arctan\frac{\sin z}{\EXP^\alpha - \cos z}\,dz\;,\quad |x|\leq 2\pi,\quad \alpha>0\;.
\end{equation}
This classical limit corresponds to $Q_{3,\delta=0}$ with $\eta_0=0$.

\subsection{A new trigonometric model with an infinite number of spin states}
There exists another trigonometric limit of the Kashiwara-Miwa
model. Consider the limit of the weights  \eqref{KM-def} when
\beq
\tau\to0, \quad N\to\infty, \quad\alpha\to0,\label{tau0}
\eeq
but the variables
\beq
\tilde{\alpha}=-\alpha/\tau,\qquad
\tcpar=-\frac{\pi}{N\tau},\qquad \re \tcpar>0, \label{tilda-def}
\eeq
are kept fixed. It is convenient to define
\beq
 w=\exp(i\tilde{\alpha}),\qquad \tilde{q}=\exp(2i{\tcpar}),\qquad |\tilde{q}|<1\ ,\label{wq-def1}
\eeq
which are exactly the same variables as in \eqref{wq-def}.
Introduce standard notations for $q$-products
\beq
(a;q)_\infty=\prod_{k=0}^\infty (1-a\,q^{k}),\quad (a;q)_n=\frac{(a;q)_\infty}
{(aq^n;q)_\infty},
\quad (a_1,a_2,\ldots,a_k;q)_n=\prod_{j=1}^k
(a_j;q)_n\ .
\eeq
Making the Jacobi imaginary transformation  in \eqref{rt-def} and
\eqref{KM-def},
dropping off associated exponentials of quadratic forms in the
spins $a,b$ (since they cancel out from the star-triangle relation)
and taking the limit \eqref{tau0}, \eqref{tilda-def},
one obtains
\beq
W(\tilde{\alpha}\,|\,a,b)=r(w\,|\,a-b)\, t(w\,|\,a+b+\zeta),\qquad
S(a)={\textstyle\hf}\ds\big(\tilde{q}^{a+\hf\zeta}+\tilde{q}^{-a-\hf\zeta}\big),
\label{trig2-def}
\eeq
where
\beq
r(w\,|\,  n)=w^{n}\,\frac{\big(\tilde{q}^\hf\,w^{-1}\,; \tilde{q}\big)_n}
{(\tilde{q}^\hf\,w\,; \tilde{q})_n},\qquad
t(w\,|\,  n)=w^{n}\,\frac{\big(-\tilde{q}^\hf\,w^{-1}\,; \tilde{q}\big)_n}
{(-\tilde{q}^\hf\,w\,; \tilde{q})_n}\ .
\eeq
The spins now take (infinitely many) arbitrary integer values,
\beq
a,b,c,\ldots \in {\mathbb Z}\ .
\eeq
The model contains two parameters: an integer $\zeta$ and a complex
parameter $\tilde{q}$,
\beq
\zeta\in {\mathbb Z},\qquad \tilde{q}\in {\mathbb C},\quad |\tilde{q}|<1\ .
\eeq
Note that by a uniform shift of all spins the
parameter $\zeta$ can be reduced to $\zeta=0$ or $\zeta=1$.
The reflection symmetry of the weights \eqref{sym2} remains intact.
The weights \eqref{trig2-def} are real
and positive when $\tilde{q}$ and $w$ are real and $1>w>\tilde{q}^{\hf}$.

The star-triangle relation for this trigonometric model reduces 
to the summation formula \eqref{hyper} for a particular 
balanced very-well-poised ${}_8\psi_8$ series, defined in \eqref{psi88}. The 
summation formula \eqref{hyper} appears to be new, see  
Appendix~\ref{app:balanced} for further details. 

The factor \eqref{fpq-def}
simplifies to
\beq
f(\tilde{\alpha})=\zsite\frac{(\tilde{q}\, w^2;\tilde{q}^2)_\infty}
{(\tilde{q}^2w^{-2}\,;\tilde{q}^2)_\infty},\qquad
\zsite=\frac{(\tilde{q}^2\,;\tilde{q}^2)_\infty}
{(\tilde{q}\,;\tilde{q}^2)_\infty},\qquad w=e^{i\tilde{\alpha}}\;,
\eeq
while \eqref{z-exp} reduces to
\begin{equation}
\kappa_e(\tilde{\alpha})\;=\;{\mathsf z}_1(w),\label{z-inf}
\end{equation}
where ${\mathsf z}_1(w)$ is defined \eqref{z12-def} (the exponential
factor from \eqref{z-exp} is absent in \eqref{z-inf} since it was
removed from the weights \eqref{trig2-def}).
Interestingly, the single-edge partition function \eqref{z-inf}
essentially coincides with that of the Ising model in \eqref{z-ising}.
Indeed, apart from the trivial exponent which was absorbed into the
normalization of weights, the two expressions differ by a rather
simple factor ${\mathsf z}_2(w)$, given by \eqref{z12-def}.

\subsubsection{Classical limit of the trigonometric model}

The classical limit corresponds to $\tilde{q}\to 1$,
\beq
\tilde{q}=e^{-\hbar}\;,\quad x=\hbar \sigma\;.
\eeq
Assuming $\tilde\alpha$ to be real and using convention (\ref{w-i-class}), one obtains the following two-point Lagrangian:
\begin{equation}
\begin{array}{l}
\ds {\mathcal L}(\tilde\alpha\,|\,x,y)\;=\;\ii \int_0^{x-y}\;\log\frac{\sinh(\hf
(z-\ii\tilde\alpha))}{\sinh( \hf (z+\ii\tilde\alpha))}\;dz\;+\;
\ii \int_0^{x+y}\;\log\frac{\cosh(\hf (z-\ii\tilde\alpha))}{\cosh( \hf
(z+\ii\tilde\alpha))}\;dz\\
[7mm]
\ds \phantom{xxxxxxxxx} \;+\;\int_0^{\tilde{\alpha}} \log |2\sin z| \,dz\;.
\end{array}
\end{equation}
This Lagrangian correspods to $Q_{3,\delta=1}$ with $\eta_0=0$.

\subsection{Gamma-function limit}\label{sec:gammamodels}

The Gamma-function limit of the hyperbolic star-triangle relations (\ref{str-hyperbolic}) for the Boltzmann weights (\ref{Q3delta1}) corresponds to
\begin{equation}\label{ratlim}
x_i\to bx_i\;,\quad i=0,1,3\;,\quad x_2\to bx_2+\ii \eta\;,\quad \alpha_i\to b\alpha_i\;,
\end{equation}
and $b^2=\ii\epsilon$, $\epsilon\to 0$. Let
\begin{equation}\label{Q2}
\begin{array}{c}
\ds {W}(\alpha\,|\,x,y)=\frac{\Gamma(\ii(x-y)-\alpha)}{\Gamma(\ii(x-y)+\alpha)}\frac{\Gamma(\ii(x+y)-\alpha)}{\Gamma(\ii(x+y)+\alpha)}\;, \\[0.5cm]
\overline{W}(\alpha\,|\,x,y)=\Gamma(\alpha+\ii(x-y))\,\Gamma(\alpha-\ii(x-y))\,\Gamma(\alpha+\ii(x+y))\,\Gamma(\alpha-\ii(x+y))\;,
\end{array}
\end{equation}
and 
\begin{equation}
{S}(x)=\frac{1}{\pi} x\sinh(2\pi x) = \frac{1}{2\, \Gamma(2\ii x)\,\Gamma(-2\ii x)}\;,
\end{equation}
where $x,y\in\mathbb{R}$, and $0<\alpha<\pi$.  The star-triangle equation is
\begin{equation}\label{str-Q2}
\begin{array}{c}
\ds\int_{-\infty}^{\infty} dx \,{S}(x) \,\overline{W}(\alpha_1\,|\,x_1,x)\, {W}(\alpha_1+\alpha_3\,|\,x_2,x) \,\overline{W}(\alpha_3\,|\,x,x_3) \\[0.3cm]
\qquad\qquad\qquad\qquad= R\,{W}(\alpha_1\,|\,x_2,x_3)\,\overline{W}(\alpha_1+\alpha_3\,|\,x_1,x_3)\, {W}(\alpha_3\,|\,x_2,x_1)\;,
\end{array}
\end{equation}
where
\begin{equation}\label{R-Q2}
R=2\pi\,\frac{\Gamma(2\alpha_1)\Gamma(2\alpha_2)}{\Gamma(2(\alpha_1+\alpha_2))}\;.
\end{equation}
For fixed choices of $x_1,x_2,x_3,\alpha_1,\alpha_3$, the integrand in \eqref{str-Q2} is an even function with respect to the integration variable $x$.  Note that in this case the star-triangle relation obtained by reversing the orientation of rapidity lines (see Section \ref{str-sect}) is unfortunately not satisfied.  

There are some differences between this model and the models of the previous subsections which are worth mentioning.  Firstly one of the weights in \eqref{Q2} is symmetric in the spins $\overline{W}(\alpha\,|\,x,y)=\overline{W}(\alpha\,|\,y,x)$, while the other one is not ${W}(\alpha\,|\,x,y)\neq {W}(\alpha\,|\,y,x)$.  In general this non-symmetric weight has a non-vanishing imaginary component $\mbox{Im}(W(\alpha\,|\,x,y))\neq0$, and thus the underlying edge-interaction model is non physical.  Also in this case the weights do not possess the simple crossing symmetry property defined in Section \ref{str-sect}, and consequently the special properties of the canonical normalisation defined in Section \ref{canonical} will not apply here.

The same limiting procedure \eqref{ratlim}, applied to the Boltzmann weights (\ref{FV}), gives the star-triangle equation (\ref{str-Q2}) with 
\begin{equation}\label{Q1d1}
\begin{array}{c}
\ds{W}(\alpha\,|\,x,y)=\frac{\Gamma(\ii(x-y)-\alpha)}{\Gamma(\ii(x-y)+\alpha)}\;,\quad {S}(x)=1\;, \\[0.6cm]
\ds\quad\overline{W}(\alpha\,|\,x,y)=\Gamma(\alpha+\ii(x-y))\Gamma(\alpha-\ii(x-y))\;,
\end{array}
\end{equation}
and $R$ given by (\ref{R-Q2}).  Similar to the preceding model, the spins take values $x,y\in\mathbb{R}$ and the spectral parameter is restricted to $0<\alpha<\pi$.  In this case, one Boltzmann weight is symmetric $\overline{W}(\alpha\,|\,x,y)=\overline{W}(\alpha\,|\,y,x)$, but the other Boltzmann weight satisfies ${W}(\alpha\,|\,y,x)= {W}^*(\alpha\,|\,x,y)$, where $^*$ denotes the complex conjugate.  In general $W(\alpha\,|\,x,y)$ has non-vanishing imaginary component $\mbox{Im}(W(\alpha\,|\,x,y))\neq0$, thus the model is non-physical.

In this case, because of the spin reflection symmetries satisfied by the Boltzmann weights, the star-triangle relation with rapidity lines reversed
\begin{equation}\label{str-Q1d1}
\begin{array}{c}
\ds\int_{-\infty}^{\infty} dx \,{S}(x) \,\overline{W}(\alpha_1\,|\,x,x_1)\, {W}(\alpha_1+\alpha_3\,|\,x,x_2) \,\overline{W}(\alpha_3\,|\,x_3,x) \\[0.3cm]
\qquad\qquad\qquad\qquad= R\,{W}(\alpha_1\,|\,x_3,x_2)\,\overline{W}(\alpha_1+\alpha_3\,|\,x_3,x_1)\, {W}(\alpha_3\,|\,x_1,x_2)\;,
\end{array}
\end{equation}
is satisfied along with \eqref{str-Q2}.  However one may restrict consideration to \eqref{str-Q2} since both of these star-triangle relations give the same three-leg forms in the quasi-classical expansion.


\subsubsection{Classical limit of the Gamma function models}
Here there is an extra complication not present in the other models of this paper.  In the case of all other models here, one may keep the unscaled spins $x_1,x_2,x_3$ and spectral parameters $\alpha_1,\alpha_3$ in the physical regime, and there is a unique maximum of the action functional $\mathcal{A}$, around which one makes the quasi-classical expansion.

For the Gamma function models one doesn't have a physical regime (one of the weights has non-vanishing imaginary component).  As discussed in Section \ref{sec:qcl}, the classical equation of motion is evaluated on solutions of the additive three-leg form of one of the $Q$-type equations of the ABS list.  However for $Q_{1,\delta=1}$ and $Q_2$, one does not have a physical regime where solutions of the additive three leg form encompass the solutions of the associated quad equations.

The classical limit of the model is just given by the Stirling approximation of Gamma-functions. The two-point Lagrangians coming from \eqref{str-Q2} are
\begin{align}
\overline{\mathcal{L}}(\alpha\,|\,x,y)= 2\int_0^{x-y}\arctan\frac{z}{\alpha}\;dz + 2\int_0^{x+y}\arctan\frac{z}{\alpha}\;dz  - 3\alpha\log|\alpha|\;,
\end{align}
and
\begin{align}
\mathcal{L}(\alpha\,|\,x,y)=-\overline{\mathcal{L}}(\alpha\,|,x,y) +2\pi |x|\;.
\end{align}
Note that $\mathcal{L}$ is not symmetric, the correct ordering of spin arguments follow from (\ref{str-Q2}). The Lagrangians are canonically normalized, coefficient (\ref{R-Q2}) is taken into account.
The one-point Lagrangian is
\begin{align}
\mathcal{C}(x)=2\pi |x|\;.
\end{align}
The classical limit of this model thus corresponds to $Q_2$ with $\eta_0=0$. Note that contrary to all other classical models with $\eta_0=0$, $\mathcal{L}$ is not symmetric, relation  (\ref{overL}) does not hold and the one-point Lagrangian is not zero.

The two-point Lagrangians coming from \eqref{str-Q1d1} is
\begin{equation}
\mathcal{L}(\alpha\,|\,x,y)=\pi |x|  - 2\int_0^{x-y}\arctan\frac{z}{\alpha}\;dz  + \alpha\log|\alpha|\;,
\end{equation}
relation (\ref{overL}) holds and one-point Lagrangian is zero. The classical limit of this model thus corresponds to $Q_{1,\delta=1}$ with $\eta_0=0$.

\subsection{Zamolodchikov's ``fishing net'' model}\label{sec:fishnet}

In 1980 Zamolodchikov \cite{Zam-fish}
obtained a solution of the star-triangle relation
motivated by considerations of ``fishing-net'' diagrams in quantum
field theory. The model has continuous spins, which are
vectors in $D$-dimensional Euclidean
space $x_1,x_2,\ldots \in {\mathbb R}^D$, where $D\ge1$.
The canonically normalized weights read \cite{Zam-fish}
\begin{equation}
\mathcal{W}(\alpha\,|\,x_1,x_2)=A(\alpha)\,
\big|x_1-x_2\big|^{-{D\alpha}/{\pi}}\;,\qquad
\mathcal{S}(x)=1,\qquad
x,x_1,x_2\in{\mathbb R}^D\;,\label{W-fish}
\end{equation}
where $|x_1-x_2|$ is the Euclidean distance between the points $x_1,x_2\in
{\mathbb R}^D$ and the crossing parameter is
\beq
\cpar=\pi\ .
\eeq
The normalization coefficient in \eqref{W-fish} is given by
\begin{equation}
A(\alpha)=
\frac{\pi^{-D\alpha/2\pi}\,\Gamma(D/2)}{\Gamma(D/2-D\alpha/2\pi)}
\prod_{\ell=1}^{\infty} \frac{\Gamma(D\ell
-D/2+D\alpha/2\pi)}{\Gamma(D\ell+D/2-D\alpha/2\pi)}
\frac{\Gamma(D\ell-D\alpha/2\pi)}{\Gamma(D\ell+D\alpha/2\pi)}
\frac{\Gamma(D\ell+D/2)}{\Gamma(D\ell-D/2)}\ ,
\end{equation}
for any values of $D\ge 1$. The last formula simplifies for even
$D$'s,
\begin{equation}
A(\alpha)=\left(\frac{D}{\pi}\right)^{D\alpha/2\pi}\;
\prod_{n=0}^{D/2-1}
\frac{\Gamma(1/2+n/D+\alpha/2\pi)}{\Gamma(1/2+n/D-\alpha/2\pi)},
\qquad D=\mbox{even} \ .
\end{equation}
The function $A(\t)$ is a ``minimal solution'' (in the sense defined
in Section \ref{sec:kmmodel}) of the functional equations
\eqref{func2}, which in this case read
\beq
A(\t)\,A(-\t)=1, \qquad A(\t)/A(\pi-\t)=\pi^{D(\pi-2\alpha)/2\pi}\frac{\Gamma(D\alpha/2\pi)}{\Gamma(D(\pi-\alpha)/2\pi)}\;.
\eeq
The boundary conditions read
\begin{equation}
\mathcal{W}(\t\,|\,x_1,x_2)=1+O(\t)\ ,  \qquad
\mathcal{W}(\pi-\t\,|\,x_1,x_2)\;=\; \delta^D(x_1-x_2)
\;+\; O(\t)\ ,\qquad \t\to 0\;,
\end{equation}
where $\delta^D(x)$ is the $D$-dimensional $\delta$-function. The
inversion relations read
\beq\begin{array}{l}
\ \ \ \ \ \ \ \ \ \ \ \ \ \ \ \ \ \mathcal{W}(\t\,|\,x_1,x_2)
\mathcal{W}(-\t\,|\,x_1,x_2)=1,\\[.4cm]
\ds\lim_{\varepsilon\to 0^+}\int  d^Dx\;
\mathcal{W}(\pi-\ii t-\varepsilon\,|\,x_1,x)\, \mathcal{W}(\pi+\ii t
-\varepsilon\,|x,x_2)=\delta^D(x_1-x_2)\;,
\end{array}\label{inv-cont}
\eeq
and $t$ is real and the integral is taken over the whole ${\mathbb
  R}^D$.
The self-duality relation reads
\begin{equation}
\mathcal{W}(\pi-\t\,|\,x_1,x_2)\;=\;\int d^Dk\
\EXP^{2\pi\ii\, k\,(x_1-x_2)}
\ \mathcal{W}(\t\,|\,k,0)\;.\label{fv-dual}
\end{equation}
The weights \eqref{W-fish} satisfy the star-triangle relation
\eqref{str-def} where the sum is replaced by the $D$-dimensional
integral over ${\mathbb R}^D$ and the parameter $\cpar=\pi$. The
coefficient ${\mathcal R}_{123}=1$, because of the canonical normalization
of weights. The star-triangle relation for this model
\begin{equation}
\begin{array}{l}
\ds\int d^Dx\,\mathcal{W}(\alpha\,|\,x,x_1)\,\mathcal{W}(\pi-\alpha-\beta\,|\,x,x_2)\,\mathcal{W}(\beta\,|\,x,x_3) \\[0.3cm]
\ds=\mathcal{W}(\pi-\alpha\,|\,x_2,x_3)\,\mathcal{W}(\alpha+\beta\,|\,x_1,x_3)\,\mathcal{W}(\pi-\beta\,|\,x_2,x_1)\;,
\end{array}
\end{equation}
follows from Symanzik's result \cite{Symanzik}.

\subsubsection{Classical limit of the fishing net model}

Consider the star-triangle equation for the fishing-net model in
the limit $D\to\infty$. External ``spins'' $x_i$ form a
three-dimensional subspace $\mathbb{E}^{d}\in\mathbb{E}^D$, where
$d=3$ but in fact we use only the condition $d\ll D$. The central spin
$x_0$ can be decomposed as
\begin{equation}
x_0=\overline{x}_0+y\;,\quad \overline{x}_0\in\mathbb{E}^d\;,\quad y\in\mathbb{E}^{D-d}\;.
\end{equation}
The saddle point of the limit $D\to\infty$ gives two equations for $y$ and $\overline{x}_0$:
the scalar equation for $y^2$,
\begin{equation}\label{z-y}
-\frac{\eta}{y^2}\;+\;\sum_{i=1}^3 \frac{\alpha_i}{(\overline{x}_0-x_i)^2+y^2}\;=\;0\;,
\end{equation}
and equation for $\overline{x}_0$ in $\mathbb{E}^d$,
\begin{equation}\label{z-x0}
\sum_{i=1}^3 \alpha_i\frac{\overline{x}_0-x_i}{(\overline{x}_0-x_i)^2+y^2}\;=\;0\;.
\end{equation}
Here
\begin{equation}
\alpha_1+\alpha_2+\alpha_3=2\eta\;,
\end{equation}
and the natural scale is $\eta=\pi$. Equations (\ref{z-y}) and (\ref{z-x0}) have the \emph{rational} solution on which the other three-legs equations are satisfied:
\begin{equation}\label{z-other}
\alpha_i\frac{x_i-\overline{x}_0}{(x_i-\overline{x}_0)^2+y^2}\;=\;
(\eta-\alpha_j)\frac{x_i-x_k}{(x_i-x_k)^2}+
(\eta-\alpha_k)\frac{x_i-x_j}{(x_i-x_j)^2}\;,\quad i,j,k=\textrm{perm}(1,2,3)\;.
\end{equation}
The crossing parameter $\eta$ is free in equations (\ref{z-y}) and
(\ref{z-x0}), and in particular $y^2\sim\eta$. Thus, in the limit
$\eta\to 0$ the variable $y$ disappears and the result is the
single saddle point equation in $\mathbb{E}^d$,
\begin{equation}\label{z-x0-0}
\sum_{i=1}^3 \alpha_i\frac{x_0-x_i}{(x_0-x_i)^2}\;=\;0\;,\quad \alpha_1+\alpha_2+\alpha_3=0\;,
\end{equation}
providing all the others:
\begin{equation}\label{z-other-0}
\alpha_i\frac{x_i-x_0}{(x_i-x_0)^2}+
\alpha_j\frac{x_i-x_k}{(x_i-x_k)^2}+
\alpha_k\frac{x_i-x_j}{(x_i-x_j)^2}\;=\;0\;,\quad i,j,k=\textrm{perm}(1,2,3)\;.
\end{equation}

In the application to the ABS-type systems, equations (\ref{z-x0-0},\ref{z-other-0}) are the three-legs equations with
\begin{equation}\label{phi}
\psi(\alpha|x,y)=\alpha\frac{x-y}{(x-y)^2}\;,\quad x,y\in\mathbb{E}^d\;,\quad \eta=0\;.
\end{equation}
Equation (\ref{z-x0-0}) in ABS notations,
\begin{equation}\label{eq1}
(\alpha_1-\alpha_2)\frac{x_{12}-x}{(x_{12}-x)^2}=\alpha_1\frac{x_1-x}{(x_1-x)^2}-\alpha_2\frac{x_2-x}{(x_2-x)^2}\;,
\end{equation}
uniquely defines the point $x_{12}$ in terms of three points
$x,x_1$ and $x_2$,
\begin{equation}\label{themap}
x_{12}\;=\;x\;+\;(\alpha_1-\alpha_2)\frac{\ds
\alpha_1\frac{x_1-x}{(x_1-x)^2}-\alpha_2\frac{x_2-x}{(x_2-x)^2}}{\ds\left(
\alpha_1\frac{x_1-x}{(x_1-x)^2}-\alpha_2\frac{x_2-x}{(x_2-x)^2}
\right)^2} \;,
\end{equation}
and gives thus the example of the vector extension of $Q_{1,\delta=0}$ \cite{MR2467378}.

\section{Conclusion}

In this paper we review the exactly solved edge interaction models of
statistical mechanics and establish their connection to classical
discrete integrable evolution equations classified by Adler Bobenko
and Suris \cite{AdlerBobenkoSuris}. We only consider the case of
a single spin degree of freedom at each site of the lattice. The
Boltzmann weights for all such models can be obtained from different
particular cases of the master solution of the star-triangle relation 
\cite{Bazhanov:2010kz}. From the algebraic point of view this solution
is related \cite{Chicherin:2014dya} 
to the modular double of the Sklyanin algebra. The corresponding
classical evolution equations are denoted as $Q_4$, they are located
at the top of the ABS list \cite{AdlerBobenkoSuris}. Similarly to the
case of lattice model the simpler equations $Q_1,Q_2,Q_3$ can in
principle be obtained as
different particular cases of $Q_4$. However, since the limiting procedure is
not always very transparent, we present separate considerations of the
corresponce to lattice models for all particular cases. The main idea
of this correspondence is related to the low-temperature (or
quasi-classical) limit of the lattice model. We found that in this
limit all 
known edge interaction models always reduce to equations from the ABS
list. The correpondence is complete in the sence that for any equation
in the ABS list there is at least one lattice counterpart. Here we
only consider the $Q$-type equations and the lattice models with
crossing symmetry. Apparently, the correspondence can easily be extended
to whole ABS list, though the related lattice models are expected to
be unphysical (i.e., to have negative Boltzmann weights, similar to
the case of $Q_2$ and $Q_{1,\delta=1}$ in this paper) and,
therefore, are of a limited interest in statistical mechanics.

\section*{Acknowledgements}
The authors thank V.Adler, A.Bobenko, F.G\"unther, R.Kashaev,
V.Mangazeev, R.Boll and Yu.Suris for fruiful discussions and correspondence. 
Andrew Kels was supported by the DFG Collaborative Research Center TRR
109, “Discretization in Geometry and Dynamics”. This work was
partially supported by the Australian Research Council (ARC).

\app{List of classical star-triangle relations}

For convenience here is a list of Lagrangian functions that appear in Section \ref{sec:models} and satisfy the star-triangle relation \eqref{cstr}.
\\\\
Master solution: $Q_4$, $\eta=\frac{\pi\tau}{2\ii}$ 
\begin{align}
\begin{array}{l}
\ds\mathcal{C}(x) = (2|x|-\frac{\pi}{2})^2\,, \\[0.5cm]
\ds\mathcal{L}(\alpha\,|\,x,y)=\frac{1}{\ii}\int_0^{x-y}\log\frac{\vartheta_4(z+\ii\alpha\,|\,\tau)}{\vartheta_4(z-\ii\alpha\,|\,\tau)}\,dz + \frac{1}{\ii}\int_{\pi/2}^{x+y}\log\frac{\vartheta_4(z+\ii\alpha\,|\,\tau)}{\vartheta_4(z-\ii\alpha\,|\,\tau)}\,dz\;.
\end{array}
\end{align}
\\
Kashiwara-Miwa model: $Q_4$, $\eta=0$
\begin{equation}
\begin{array}{l}
\ds\mathcal{C}(x)=0\;,\\[5mm]
\ds \mathcal{L}(\ii\alpha\,|\,x,y) \;=\; \ii \int_0^{x-y}\;
\log\frac{\vartheta_1(\hf (z-\ii\alpha)\,|\,\tau)}{\vartheta_1(\hf
(z+\ii\alpha)\,|\,\tau)}\;dz\;+\;\ii\int_0^{x+y}\; \log\frac{\vartheta_1(\hf
(z-\ii\alpha)\,|\,\tau)}{\vartheta_1(\hf
(z+\ii\alpha)\,|\,\tau)}\;dz\\
[7mm]
\ds\phantom{xxxxxxxxxx}
+\;\int_0^\alpha \log\left|\frac{\ds\vartheta_1(\ii\frac{z}{2}\,|\,\frac{\tau}{2})}{\ds\vartheta_2(\ii\frac{z}{2}\,|\,\frac{\tau}{2})}\right| \;dz\;.
\end{array}
\end{equation}
\\
Hyperbolic limit of master solution: $Q_{3,\delta=1}$, $\eta=\frac{\pi}{2}$ 
\begin{equation}
\begin{array}{l}
\ds\mathcal{C}(x)=-2\pi|x|\,, \\[0.5cm]
\ds\mathcal{L}(\alpha\,|\,x,y) = \frac{1}{\ii} \int_0^{x-y} \log\frac{\cosh(z+\ii\alpha)}{\cosh(z-\ii\alpha)}\,dz +
\frac{1}{\ii}\int_0^{x+y} \log\frac{\cosh(z+\ii\alpha)}{\cosh(z-\ii\alpha)}\,dz + 2\int_0^{\alpha}\log2\cos(z)\,dz\;,
\end{array}
\end{equation}
\\
Trigonometric solution: $Q_{3,\delta=1}$, $\eta=0$ 
\begin{equation}
\begin{array}{l}
\ds\mathcal{C}(x)=0\;,\\[5mm]
\ds {\mathcal L}(\tilde\alpha\,|\,x,y)\;=\;\ii \int_0^{x-y}\;\log\frac{\sinh(\hf
(z-\ii\tilde\alpha))}{\sinh( \hf (z+\ii\tilde\alpha))}\;dz\;+\;
\ii \int_0^{x+y}\;\log\frac{\cosh(\hf (z-\ii\tilde\alpha))}{\cosh( \hf
(z+\ii\tilde\alpha))}\;dz\\
[7mm]
\ds \phantom{xxxxxxxxx} \;+\;\int_0^{\tilde{\alpha}} \log |2\sin z| \,dz\;.
\end{array}
\end{equation}
\\
Faddeev-Volkov model: $Q_{3,\delta=0}$, $\eta=\frac{\pi}{2}$
\begin{equation}
\begin{array}{l}
\ds\mathcal{C}(x)=0\,, \\[0.5cm]
\ds\mathcal{L}(\alpha\,|\,x,y)=\frac{1}{\ii} \int_0^{x-y} \log\frac{\cosh(z+\ii\alpha)}{\cosh(z-\ii\alpha)} dz +2\int_0^\alpha \log 2\cos(z)dz \;.
\end{array}
\end{equation}
\\
Fateev-Zamolodchikov model: $Q_{3,\delta=0}$, $\eta=0$
\begin{equation}
\begin{array}{l}
\ds \mathcal{C}(x)=0\;,\\[5mm]
\ds {\mathcal L}(\ii\alpha\,|\,x,y)\;=\;\pi|x| + \ii\int_0^{x-y} \;
\log\frac{\sinh(\hf (\alpha+\ii z))}{\sinh(\hf (\alpha-\ii z))}dz \;+\;\int_0^\alpha \log\tanh\frac{z}{2}\,dz\;.
\end{array}
\end{equation}
\\
Gamma function model: $Q_2$, $\eta=0$
\begin{equation}
\begin{array}{l}
\ds\mathcal{C}(x)=2\pi|x|\,, \\[0.5cm]
\ds\mathcal{L}(\alpha\,|\,x,y)= -2\int_0^{x-y}\arctan\frac{z}{\alpha}\;dz - 2\int_0^{x+y} \arctan\frac{z}{\alpha}\;dz +3\alpha\log|\alpha| +2\pi|x|\;,\\[0.5cm]
\ds\overline{\mathcal{L}}(\alpha\,|\,x,y) = 2\int_0^{x-y}\arctan\frac{z}{\alpha}\;dz + 2\int_0^{x+y} \arctan\frac{z}{\alpha}\;dz  -3\alpha\log|\alpha|\;.
\end{array}
\end{equation}
\\
Gamma function model: $Q_{1,\delta=1}$, $\eta=0$
\begin{equation}
\begin{array}{l}
\ds\mathcal{C}(x)=0\,, \\[0.5cm]
\ds\mathcal{L}(\alpha\,|\,x,y)=\pi |x| -2\int_0^{x-y}\arctan\frac{z}{\alpha}\;dz + \alpha\log|\alpha|\;.
\end{array}
\end{equation}
\\
$D=1$ Fishing net model: $Q_{1,\delta=0}$, $\eta=0$
\begin{equation}
\begin{array}{l}
\ds\mathcal{C}(x)=0\,, \\[0.3cm]
\ds\mathcal{L}(\alpha\,|\,x,y)=\alpha\log\left|x-y\right|-\frac{\alpha}{2}\log\left|{\alpha}\right|
\end{array}
\end{equation}

\app{Identity for the derivatives of the three-leg form}

The next-to-leading order quasi-classical expansion of the star-triangle relation, gives the following new identity involving the derivative of a three leg equation
\begin{equation}
\begin{array}{r}
\ds-\frac{\partial^2}{\partial x^2}\left(\mathcal{L}(\eta-\alpha\,|\,x,x_1)+\mathcal{L}(\alpha+\beta\,|\,x,x_2)+\mathcal{L}(\eta-\beta\,|\,x_3,x)\right)_{x=x_0} \\[0.6cm]
\ds=\left.\frac{{\mathcal{L}}^{(0)}(\eta-\alpha-\beta\,|\,x_1,x_3)}{{\mathcal{L}}^{(0)}(\alpha\,|\,x,x_1){\mathcal{L}}^{(0)}(\beta\,|\,x_3,x)}\right|_{x=x_0}\;.
\end{array}
\end{equation}
The Lagrangians $\mathcal{L}(\alpha\,|\,x,y)$ are listed in Appendix A, and $x_0$ is the unique solution to the three leg equation
\begin{equation}
\frac{\partial}{\partial x}\left(\mathcal{L}(\eta-\alpha\,|\,x,x_1)+\mathcal{L}(\alpha+\beta\,|\,x,x_2)+\mathcal{L}(\eta-\beta\,|\,x_3,x)\right)=0\,,
\end{equation}
for fixed choices of $x_1,x_2,x_3,\alpha,\beta$.  An example of the function $\mathcal{L}^{(0)}(\alpha\,|\,x,y)$ for each of the equations in the $Q$ list is given below.
\\\\
{Master solution} ($Q_4$, $\eta\neq 0$):
\begin{equation}
{\mathcal{L}}^{(0)}(\alpha\,|\,x,y)=\frac{-2\ii\, \vartheta_1(2x\,|\,\tau)\,\vartheta_1(2y\,|\,\tau)\,\vartheta_1(2\ii\alpha\,|\,\tau)\left((p^2;p^2)_\infty\right)^3p^{1/4}}{\vartheta_1(x+y+\ii\alpha\,|\,\tau)\,\vartheta_1(x-y+\ii\alpha\,|\,\tau)\,\vartheta_1(x+y-\ii\alpha\,|\,\tau)\,\vartheta_1(x-y-\ii\alpha\,|\,\tau)}\,.
\end{equation}
\\
{Hyperbolic limit of the master solution} ($Q_{3,\delta=1}$, $\eta\neq 0$):
\begin{equation}
{\mathcal{L}}^{(0)}(\alpha\,|\,x,y)=\frac{\sinh x\sinh y\sin\alpha}{2\sinh\frac{x+y+\ii\alpha}{2}\sinh\frac{x-y+\ii\alpha}{2}\sinh\frac{x+y-\ii\alpha}{2}\sinh\frac{x-y-\ii\alpha}{2}}\,.
\end{equation}
\\
{Faddeev-Volkov model} ($Q_{3,\delta=0}$, $\eta\neq 0$)
\begin{equation}
\begin{array}{l}
\ds{\mathcal{L}}^{(0)}(\alpha\,|\,x,y)=\frac{\sin\alpha}{2\sinh\frac{x-y+\ii\alpha}{2}\sinh\frac{x-y-\ii\alpha}{2}}\,.
\end{array}
\end{equation}
\\
{Gamma function model} ($Q_2$, $\eta=0$):
\begin{equation}
{\mathcal{L}}^{(0)}(\alpha\,|\,x,y)=\frac{8xy\alpha}{(x+y+\ii\alpha)(x+y-\ii\alpha)(x-y+\ii\alpha)(x-y-\ii\alpha)}\,.
\end{equation}
\\
{Gamma function model} ($Q_{1,\delta=1}$, $\eta=0$):
\begin{equation}
{\mathcal{L}}^{(0)}(\alpha\,|\,x,y)=\frac{2\alpha}{(x-y+\ii\alpha)(x-y-\ii\alpha)}\,.
\end{equation}
\\
{$d=1$ Zamolodchikov fishing net model} ($Q_{1,\delta=0}$, $\eta=0$):
\begin{equation}
{\mathcal{L}}^{(0)}(\alpha\,|\,x,y)=\frac{|\alpha|}{(x-y)^2}\,.
\end{equation}

\app{Limits of elliptic gamma-functions.}\label{phi-prop}

The elliptic gamma-function $\Phi(z)$ is defined by (\ref{Phi}). Two particular limits should be considered. The classical limit corresponds to
\begin{equation}
p=\EXP^{\ii\pi\tau'}\;\to\;1\;,\quad \tau'\to 0\;.
\end{equation}
In this limit the leading asymptotics are
\begin{equation}
\Phi(z)\;=\;\exp\left\{ -\frac{1}{\pi\tau'} \int_0^z \log\overline{\vartheta}_4(t\,|\,\tau) \,dt + {O}(1)\right\}\;.
\end{equation}
For the normalization function $K(\alpha)$ defined in \eqref{Kappa},
\begin{equation}
K(\alpha)\;=\;\exp\left\{-\frac{1}{\pi\tau'} \int_0^{2\ii\alpha} \log\overline{\vartheta}_4(t\,|\,2\tau) \,dt + {O}(1)\right\}\;.
\end{equation}
Here the Jacobi theta function $\overline{\vartheta}_4$ is given by
\begin{equation}
\overline{\vartheta}_4(z\,|\,\tau) \;=\; \prod_{n=0}^\infty (1-\EXP^{2\ii z} q^{2n+1})(1-\EXP^{-2\ii z}q^{2n+1})\;.
\end{equation}
Another limit is the hyperbolic case. Let
\begin{equation}
\tau=\ii\, \frac{b}{T}\;,\quad \tau'=\ii\,\frac{b^{-1}}{T}\;,\quad T\to \infty\;.
\end{equation}
Below we use the hyperbolic notations
\begin{equation}\label{hypnot}
\eta=\frac{1}{2}(b+b^{-1}),\qquad
q=\EXP^{\ii\pi b^2},\qquad \tilde{q}=\EXP^{-\ii\pi b^{-2}},\qquad
\overline{q}=\ii\,\exp\left(\frac{\ii\pi(b-b^{-1})}{2(b+b^{-1})}\right)\ ,
\end{equation}
\begin{equation}
(x,q)_\infty\,\be\, \prod_{k=0}^\infty (1-q^k x)\ .
\end{equation}
In this limit
\begin{equation}
\log\Phi(z)=-\ii\frac{T^2}{6}z(1-\frac{z}{\pi})(1-\frac{2z}{\pi}) + \frac{\ii}{12}(b^2+b^{-2})(\frac{\pi}{2}-z) + {O}(T^{-2})\;.
\end{equation}
This is a $\pi$-periodic function written in the interval $0<z<\pi$. The regular term here has a jump discontinuity at $z=0$, so there is another limit for small $z$:
\begin{equation}
\Phi(\frac{\pi}{T} z) = \EXP^{ -\ii \frac{\pi}{6} T z}  \phi(z)\;,
\end{equation}
where
\begin{equation}
\phi(z)=\exp\left\{\frac{1}{4} \int_{pv} \frac{\EXP^{-2\ii\sigma w}}{\sinh (bw) \sinh (b^{-1}w)} \frac{dw}{w}\right\} \;,
\end{equation}
is the symmetric quantum dilogarithm. It is related to the usual quantum dilogarithm
\begin{equation}\label{C11}
\varphi(z)=\exp\left\{\frac{1}{4} \int_{\mathbb{R}+\ii 0} \frac{\EXP^{-2\ii\sigma w}}{\sinh (bw) \sinh (b^{-1}w)} \frac{dw}{w}\right\} \;,
\end{equation}
by
\begin{equation}
\phi(z)=\EXP^{-\frac{\ii\pi}{2} z^2+\frac{\ii\pi}{12}(1-2\eta^2)}\varphi(z)\;.
\end{equation}
The normalization function $K(\alpha)$ has the limit
\begin{equation}
K(\frac{\pi}{T}\,\alpha) = \EXP^{\frac{\pi}{6} T\alpha} \kappa(\alpha)\;,
\end{equation}
where
\begin{equation}
\kappa(\alpha)=\exp\left\{\frac{1}{8}\int_{pv} \frac{\EXP^{2\alpha w}}{\sinh (bw) \sinh (b^{-1}w) \cosh ((b+b^{-1})w)} \frac{dw}{w}\right\}\;.
\end{equation}
Removal of the principal value gives
\begin{equation}
\kappa(\alpha)=\EXP^{\ii\pi\alpha^2 + \frac{\ii\pi}{24}(1-8\eta^2)}\Gfun(2\ii\alpha)\;,
\end{equation}
where
\begin{equation}
\Gfun(z)=\exp\left\{\frac{1}{8} \int_{\mathbb{R}+\ii 0} \frac{\EXP^{-2\ii zw}}{\sinh (bw) \sinh(b^{-1}w) \cosh((b+b^{-1})w)} \frac{dw}{w}\right\}\;.
\end{equation}
In the limit $b^2=\ii\epsilon$ and $\epsilon\to 0$, the leading symptotics of the above functions are
\begin{equation}
\varphi(bz+\ii\eta)\;\sim\;\frac{1}{2\pi\ii\sqrt{\epsilon}}\, \EXP^{-\pi/12\epsilon} \frac{(2\pi\epsilon)^{\ii z}}{\EXP^{\pi z}}\, \Gamma(\ii z)\;,
\end{equation}
\begin{equation}
\phi(b z + \ii \eta)\;\sim\;\frac{1}{2\pi b}  \frac{(2\pi\epsilon)^{\ii z}}{\EXP^{\pi z/2}} \,\Gamma(\ii z)\;,
\end{equation}
and
\begin{equation}
\kappa(\eta-b\alpha)\sim\frac{1}{2\pi b} \EXP^{\pi\ii \alpha}(\pi^2\epsilon^2(1-b^2))^\alpha\,\Gamma(2\alpha)\;.
\end{equation}

The classical limit for the hyperbolic functions $\phi(z)$ and $\kappa(\alpha)$ in the limit $b\to 0$ are
\begin{equation}
\phi(\frac{z}{\pi b}) = \exp\left\{ \frac{\ii}{\pi b^2} \int_0^z \log 2\cosh x dx +  {O}(b^2)\right\}\;,
\end{equation}
and
\begin{equation}
\kappa(\frac{\alpha}{\pi b}) = \exp\left\{\frac{2\ii}{\pi b^2} \int_0^{\ii\alpha} \log 2\cosh x dx + {O}(1)\right\}\;.
\end{equation}


\app{$D=1$ fishnet model limit of master solution}


The algebraic limit of the master solution is related to the Zamolodchikov fishnet model of Section \ref{sec:fishnet}, with $D=1$.  This limit was previously considered by Rains \cite{Rains,Rains2}.  Recall the master solution of the star-triangle relation of Section \ref{sec:models} in the equivalent following form
\begin{align}
\int^{2\pi}_0dx_0\,\mathcal{S}(x_0)W(\eta-\alpha_1\,|\,x_1,x_0)W(\alpha_1+\alpha_3\,|\,x_2,x_0)W(\eta-\alpha_3\,|\,x_3,x_0) \nonumber \\
=R\,W(\alpha_1\,|\,x_2,x_3)W(\eta-\alpha_1-\alpha_3\,|\,x_1,x_3)W(\alpha_3\,|\,x_2,x_1)
\end{align}
where
\begin{align}
W(\alpha\,|\,x,y)=\frac{\Phi(x-y+\ii\alpha)\Phi(x+y+\ii\alpha)}{\Phi(x-y-\ii\alpha)\Phi(x+y-\ii\alpha)}\;,
\end{align}
and
\begin{align}
\mathcal{S}(x)=\frac{(p^2;p^2)_\infty(q^2;q^2)_\infty}{4\pi}\Phi(2x-\ii\eta)\Phi(-2x-\ii\eta),\quad R&=\frac{\Phi(\ii\eta-2\ii\alpha_1)\Phi(\ii\eta-2\ii\alpha_3)}{\Phi(\ii\eta-2\ii(\alpha_1+\alpha_3))}
\end{align}
The elliptic gamma function $\Phi$ is defined in \eqref{Phi}, and $(x;p)_\infty=\prod^\infty_{n=1}(1-xp^n)$.  To obtain the algebraic limit the first step is to take the limit of the spectral parameters and elliptic nomes
\begin{align}
\alpha_i=\epsilon\alpha_i\;,\quad p=\EXP^{\ii\pi\tau}>0\;,\quad q=\EXP^{-2\epsilon}\;,\quad\eta=\epsilon-\frac{\ii\pi\tau}{2}\;,\quad\epsilon\rightarrow0\;.
\end{align}
As $\epsilon\rightarrow0$, the asymptotics of the elliptic gamma function are given by
\footnote{We understand $|\vartheta_1(x\,|\,\tau)$ as
$$
|\vartheta_1(x\,|\tau)| = 2p^{1/4}|\sin x| (p^2\EXP^{2\ii x};p^2)_\infty(p^2\EXP^{-2\ii x};p^2)_\infty(p^2;p^2)_\infty
$$
}
\begin{align}
\begin{array}{rcl}
\ds\log\frac{\Phi(x+\ii\alpha\epsilon)}{\Phi(x-\ii\alpha\epsilon)}&\!\!\!\!=\!\!\!\!&\ds-\alpha\log\frac{\vartheta_4(x\,|\,\tau)}{(p^2,p^2)_\infty}+O(\epsilon)\;, \\[0.6cm]
\ds\log\frac{\Phi(x+\ii(\eta-\alpha\epsilon))}{\Phi(x-\ii(\eta-\alpha\epsilon))}&\!\!\!\!=\!\!\!\!&\ds\frac{1}{4\epsilon}\left(\mbox{Li}_2\,(\EXP^{2\ii x})+\mbox{Li}_2\,(\EXP^{-2\ii x})\right) \\[0.4cm]
&&\ds-(1-\alpha)\log\frac{\left|\vartheta_1(x\,|\,\tau)\right|}{ p^{1/4}(p^2,p^2)_\infty}+O(\epsilon)\;,
\end{array}
\end{align}
This gives the following asymptotic of the Boltzmann weights of the master solution \eqref{q4wts}
\begin{align}
\begin{array}{rcl}
\ds\log{W}(\alpha\epsilon\,|\,x,y)&\!\!\!\!=\!\!\!\!&\ds-\alpha\log\frac{\vartheta_4(x\pm y\,|\,\tau)}{(p^2,p^2)^2_\infty}+O(\epsilon)\;, \\[0.6cm]
\ds\log{W}(\eta-\alpha\epsilon\,|\,x,y)&\!\!\!\!=\!\!\!\!&\ds\frac{1}{\epsilon}\left(\frac{\pi^2}{6}-\frac{\pi}{2} |x+y| -\frac{\pi}{2}|x-y| + x^2 + y^2)\right) \\[0.5cm]
&&\ds\quad-(1-\alpha)\log\frac{\left|\vartheta_1(x\pm y\,|\,\tau)\right|}{p^{1/2}(p^2,p^2)^2_\infty}+O(\epsilon)\;,
\end{array}
\end{align}
and
\begin{align}
\begin{array}{rcl}
\ds\log\mathcal{S}(x)&\!\!\!\!=\!\!\!\!&\ds-\frac{1}{\epsilon}\left(\frac{\pi^2}{8}+2x^2-\pi|x|\right)-\frac{1}{2}\log(8\pi\epsilon)+\log p^{-1/4}\left|\vartheta_1(2x\,|\,\tau)\right|+O(\epsilon)\;, \\[0.6cm]
\ds\log{R}&\!\!\!\!=\!\!\!\!&\ds\frac{\pi^2}{24\epsilon}-\frac{1}{2}\log(8\pi\epsilon)-\log(p^2,p^2)_\infty+\log\frac{\Gamma(\alpha_1)\,\Gamma(\alpha_3)}{\Gamma(\alpha_1+\alpha_3)}+O(\epsilon)\;.
\end{array}
\end{align}
Both sides of the star-triangle relation are  exponentially suppressed by $\EXP^{-\pi/\epsilon}$ outside the region 
\begin{align}
x_0\in B\;\equiv\; [-\mbox{max}(|x_1|,|x_3|),-\mbox{min}(|x_1|,|x_3|)]\;\cup\; [\mbox{min}(|x_1|,|x_3|),\mbox{max}(|x_1|,|x_3|)]\;.
\end{align}
One then finds the following Boltzmann weights
\begin{align}
\label{Jacobistr}
\begin{array}{rl}
\ds{W}(\alpha\,|\,x,y)=\vartheta_4(x\pm y\,|\,\tau)^{-\alpha}\;,\quad&\ds\overline{{W}}(\alpha\,|\,x,y)=\left|\vartheta_1(x\pm y\,|\,\tau)\right|^{-(1-\alpha)}\;, \\[0.3cm]
\ds\mathcal{S}(x)=\left|\eta^3(\tau)\vartheta_1(2x\,|\,\tau)\right|\;,\quad&\ds{R}=\frac{\Gamma(\alpha_1)\,\Gamma(\alpha_3)}{\Gamma(\alpha_1+\alpha_3)}\;,
\end{array}
\end{align}
where $\eta(\tau)=p^{1/2}(p^2;p^2)_\infty$, that satisfy the star-triangle relation
\begin{align}
\int_{B}dx_0\,\mathcal{S}(x_0)\overline{{W}}(\alpha_1\,|\,x_1,x_0){W}(\alpha_1+\alpha_3\,|\,x_2,x_0)\overline{{W}}(\alpha_3\,|\,x_3,x_0) \\
={R}{W}(\alpha_1\,|\,x_2,x_3)\overline{{W}}(\alpha_1+\alpha_3\,|\,x_1,x_3){W}(\alpha_3\,|\,x_2,x_1)\;,
\end{align}
The second step is to substitute the following function
\begin{align}
y_i = \frac{\vartheta_1(x_i\,|\,\tau)^2}{\vartheta_4(x_i\,|\,\tau)^2}\;,\quad i=0,1,3\;,\quad 
y_2 = \frac{\vartheta_4(x_2\,|\,\tau)^2}{\vartheta_1(x_2\,|\,\tau)^2}\;,\quad \frac{\vartheta_4(\tau)^2\mathcal{S}(x_0)}{\vartheta_4(x_0\,|\,\tau)^4}dx_0=\frac{1}{2}dy_0\;,
\end{align}
into the star-triangle relation \eqref{Jacobistr}, and after a change of variables one finds the Boltzmann weights
\begin{align}
\label{quan1d0}
{W}(\alpha\,|\,x,y)=\left|x-y\right|^{-\alpha}\;,\quad{R}=\frac{\Gamma(\alpha_1)\Gamma(\alpha_3)}{\Gamma(\alpha_1+\alpha_3)}\;.
\end{align}
These Boltzmann weights satisfy the star-triangle relation 
\begin{align}
\label{Selberg}
\int_{x_1}^{x_3}dx_0\,{W}(\eta-\alpha_1\,|\,x_1,x_0){W}(\alpha_1+\alpha_3\,|\,x_2,x_0){W}(\eta-\alpha_3\,|\,x_3,x_0) \\
={R}{W}(\alpha_1\,|\,x_2,x_3){W}(\eta-\alpha_1-\alpha_3\,|\,x_1,x_3){W}(\alpha_3\,|\,x_2,x_1)\;
\end{align}
with $\eta=1$ and assumption $x_1<x_0<x_3<x_2$, and is equivalent to the Selberg integral \cite{Rains,Rains2}.

The star-triangle relation \eqref{quan1d0} is contained in the $D=1$ Zamolodchikov fishing net model, the latter being given in terms of the same weights ${W}$ in \eqref{quan1d0}\footnote{These weights differ from \eqref{W-fish} only by a trivial rescaling $\alpha\leftrightarrow\pi\alpha$}, only  
${R}$ for Zamolodchikov fishing net is given by
\begin{align}
\ds{R}=\sqrt{\pi}\,\frac{\Gamma(\frac{\alpha_1}{2})\,\Gamma(\frac{1-(\alpha_1+\alpha_3)}{2})\,\Gamma(\frac{\alpha_3}{2})}{\Gamma(\frac{1-\alpha_1}{2})\,\Gamma(\frac{\alpha_1+\alpha_3}{2})\,\Gamma(\frac{1-\alpha_3}{2})}\;,
\end{align}
and the integration in \eqref{Selberg} is taken over the real axis.  This form of the star-triangle relation is more satisfactory from physical considerations, as the value of different spins should be independent of one another.

\app{Analytic propreties of $\varphi$ and $\Gfun$}

\noindent
{\em The function $\varphi(z)$.}
This function is defined by the integral \eqref{C11}. It has
the following properties (with the notations defined in \eqref{hypnot})  
\begin{enumerate}[(a)]
\item
{\em Simple poles and zeros}
\begin{equation}\label{poles}
\begin{array}{l}
\textrm{poles of } \varphi(z)\;=\;\left\{
\ii\crs + \ii m b + \ii n b^{-1}\;,\;\;\;m,n\in\mathbb{Z}_{\geq 0}\right\}\;,\\
\textrm{zeros of } \varphi(z)\;=\;\left\{ -(\ii \crs + \ii m b +
\ii n b^{-1})\;,\;\;\;m,n\in\mathbb{Z}_{\geq 0}\right\}\;.
\end{array}
\end{equation}
\item
{\em Functional relations}
\begin{equation}\label{difference}
\varphi(z)\varphi(-z)\;=\;\EXP^{\ii\pi z^2 - \ii\pi
(1-2\crs^2)/6}\;,\qquad  \frac{\varphi(z-\ii b^{\pm 1}/2)}
{\varphi(z+\ii b^{\pm 1}/2)}\;=\; \left(1\,+\,\EXP^{2\pi z
b^{\pm 1}}\right)\;.
\end{equation}
\item
{\em Asymptotics}
\begin{equation}
\varphi(z)\;\simeq\;1,\qquad\Re (z)\to
-\infty;\qquad\varphi(z)\;\simeq\;\EXP^{\ii\pi
z^2-\ii\pi(1-2\crs^2)/6}\qquad \Re (z)\to
+\infty\;.
\end{equation}
where $\Im (z)$ is kept finite.
\item
{\em Product representation}
\begin{equation}
\varphi(z)=\frac{(-q\,e^{2\pi z\,b}\,;\ q^2)_\infty}
{( -\tilde q\,e^{2\pi z\, b^{-1}};\tilde  q^{\,2})_\infty},\qquad \Im\, b^2>0\;.
\end{equation}
\item
{\em Pentagon relation.} The function $\varphi(z)$ satisfy the
following operator identity \cite{Faddeev:1994}
\begin{equation}
\varphi(\mathsf{P}) \varphi(\mathsf{X}) \;=\;
\varphi(\mathsf{X}) \varphi(\mathsf{P}+\mathsf{X})
\varphi(\mathsf{P})\;,\quad
[\mathsf{P},\mathsf{X}]\;=\;\frac{1}{2\pi\ii}\;.\label{pent}
\end{equation}
where $[\ ,\ ]$ denotes the commutator. It can be re-written in the matrix
form \cite{Ponsot:2001,Kas01}
\begin{equation}\label{rstr}
\ds \int_{\mathbb{R}}\; \frac{\varphi(x+u)}{\varphi(x+v)}\;
\EXP^{2\pi\ii wx}\;dx\;=\; \EXP^{\ii\pi(1+4\crs^2)/12\ -2\pi\ii
w(v+\ii\crs)}\
\frac{\varphi(u-v-\ii\crs)\varphi(w+\ii\crs)}{\varphi(u-v+w-\ii\crs)}\;,
\end{equation}
where the wedge of poles of $\varphi(x+u)$ must lie in the upper
half-plane, the wedge of zeros of $\varphi(x+v)$ must lie in the
down half-plane, and the integrand must decay when
$x\to\pm\infty$.
\end{enumerate}

\noindent
{\em The function $\Gfun(z)$.} This function is defined by the integral
\eqref{fvkappa}. It has the following properties.
\begin{enumerate}[(i)]
\item
{\em Simple poles and zeros}
\begin{equation}\label{G-poles}
\begin{array}{l}
\ds\textrm{poles of $\Gfun(z)$} \;=\; \left\{ 2\ii\crs  + \ii m b
+ \ii n b^{-1}\;, \;\;\; m,n\in \mathbb{Z}_{\geq 0}\;,\;\;\;
m+n-|m-n| = 0 \mod 4 \right\}\;,\\
\ds\textrm{zeros of $\Gfun(z)$} \;=\; \left\{ -(2\ii\crs + \ii m b
+ \ii n b^{-1}),\;\; m,n\in \mathbb{Z}_{\geq 0},\;\;
m+n-|m-n| = 0 \mod 4\right\}.
\end{array}
\end{equation}
\item
{\em Functional relations}
\begin{equation}
\Gfun(z)\;\Gfun(-z)\;=\;\EXP^{\ii\pi z^2/2 - \ii \pi
(1-8\crs^2)/12}\;,\quad \Gfun(z+\ii\crs) \Gfun(z-\ii\crs) \;=\;
\varphi(z)\;.
\end{equation}
\item
{\em Asymptotics}
\begin{equation}
\Gfun(z)\simeq 1,\qquad z\to -\infty;\qquad
\Gfun(z)\simeq \EXP^{\ii\pi z^2/2 -
\ii\pi(1-8\crs^2)/12},\qquad z\to +\infty\ .
\end{equation}
where $\Im (z)$ is kept finite.
\item
{\em Product representation}
\begin{equation}
\Phi(z)=\frac{(q^2 e^{2\pi z\,b};q^4)_\infty}
{(\tilde{q}^{\,2}e^{2\pi z\, b^{-1}};\tilde{q}^{\,4})_\infty}
\frac{\ds(-\overline{q} e^{{\pi z}/(2\lambda)};
\overline{q}^{\,2})_\infty}
{\ds(\overline{q} \,e^{{\pi z}/(2\lambda};\overline{q}^{\,2})_\infty},
\qquad \Im\, b^2>0\ .
\end{equation}
\end{enumerate}

\app{\label{app:balanced}
Summation formula for a balanced
very-well-poised ${}_8\psi_8$ series.}

The weights \eqref{trig2-def}
satisfy the star-triangle relation \eqref{str-def}. Introducing the variables
$v_i=e^{i\tilde{\alpha}_i}$, $i=1,2,3$, \  and denoting the integer spins
$(a,b,c)$ as $(\sigma_1,\sigma_2,\sigma_3)$, and for shortness using $q$ instead of $\tilde{q}$, one can rewrite \eqref{str-def}
as an identity for basic hypergeometric series,
\beq
\begin{array}{c}
\ds
\sum_{n=-\infty}^{\infty}
\frac{\ds q^n+q^{3n+\zeta}}{\ds 1+q^\zeta}\,
\prod_{j=1}^3
\frac{\ds (q^{-\sigma_j}v_j,-q^{\sigma_j+\zeta}v_j;q)_n}
{\ds (q^{1-\sigma_j}/v_j,-q^{1+\sigma_j+\zeta}/v_j;q)_n}
=\ds \left(\frac{2}{1+q^\zeta}\right)
\frac{\ds (q^2,\,qv_1^2,\,qv_2^2,\,qv_3^2;\,q^2)_\infty}
{\ds(q,\,q^2/v_1^2,\,q^2/v_2^2,\,q^2/v_3^2;\,q^2)_\infty}\\[1.0cm]\ds
\times \Bigg[\,\prod_{i=1}^3
\frac{\ds (q/v_i\,;q)_{\sigma_i}\,(-q/v_i\,;q)_{\sigma_i+\zeta}}
{\ds (v_i\,;q)_{\sigma_i}\,(-v_i\,;q)_{\sigma_i+\zeta}}\,\Bigg]
\ \left[\,\prod_{i=1}^3
\frac{\ds v_i^{-2\sigma_k}\,(q^\hf/v_i;\,q)_{\sigma_j-\sigma_k}
\,(-q^\hf/ v_i;\,q)_{\sigma_j+\sigma_k+\zeta}}
{\ds (q^\hf v_i;\,q)_{\sigma_j-\sigma_k}
\,(-q^\hf v_i;\,q)_{\sigma_j+\sigma_k+\zeta}}\right]\ ,
\end{array}\label{hyper}
\eeq
where $(i,j,k)=\mbox{cycle}(1,2,3)$ in the second product is a
cyclic permutation of $(1,2,3)$. The quantities
$\sigma_1,\sigma_2,\sigma_3, \zeta\in{\mathbb Z}$ are arbitrary
integers, the quantities $q,v_1$, $v_2$ are indeterminates and
$v_3\equiv q^\hf/v_1v_2$.

Several remarks are in order. First,
the identity \eqref{hyper} is unvariant under simultaneous shifts
\beq
\zeta\to\zeta-2d,\qquad \sigma_i\to\sigma_i+d,\qquad d\in{\mathbb Z}
\qquad i=1,2,3\ .
\eeq
This is especially obvious in the original star-triangle form of this
relation.
Thus, without loss of generality, one can assume that $\zeta=0$ or $\zeta=1$.
Second, with the standard notation \cite{gasper} for the bilateral basic
hypergemetric series
\beq
 \;_k\psi_k \left[\begin{matrix} a_1 & a_2 &
\ldots & a_k \\ b_1 & b_2 & \ldots & b_k \end{matrix} \Big|\, q;z \right]
= \sum_{n=-\infty}^\infty \frac {(a_1, a_2, \ldots, a_k;q)_n}
{(b_1, b_2, \ldots, b_k;q)_n}\  z^n\ ,
\eeq
the sum in the LHS of \eqref{hyper} can be written as
\beq
\;_8\psi_8\Bigg[
\begin{array}{llllllll}
q\,z^\hf,&\!\!\!- q\,z^\hf,&\!\!\!
v_1/s_1,&\!\!\! z \,s_1\,v_1,&\!\!\!
v_2/s_2,&\!\!\! z \,s_2\,v_2,&\!\!\!
q^{\hf}/s_3\,v_1\,v_2,&\!\!\!q^{\hf}\,z \,s_3/v_1\,v_2\\[.2cm]
z^\hf,&\!\!\!-z^\hf,&\!\!\!
q\,z\,s_1/v_1,&\!\!\!q/s_1\, v_1,&\!\!\!
q\,z\,s_2/v_2,&\!\!\!q/s_2\, v_2,&\!\!\!
q^\hf\, z\,s_3 \,v_1 \,v_2,&\!\!\! q^{\hf}\,v_1\,v_2/s_3
\end{array}\Big|\,q;q\Bigg],\label{psi88}
\eeq
where we have denoted
\beq
z^\hf={\mathsf i\,} q^{\hf\zeta},\quad s_1=q^{\sigma_1},
\quad s_2=q^{\sigma_2},\quad s_3=q^{\sigma_3},\quad \sigma_1,\sigma_2,
\sigma_3,\zeta\in{\mathbb Z}\ .
\eeq
Examining relationships between arguments in \eqref{psi88} one concludes that
\eqref{hyper} is a summation formula for a particular\footnote{
A bilateral basic hypergeometric
series ${}_k\psi_k$ is well-poised if $a_1b_1 = a_2b_2= \cdots =
a_kb_k$ and very-well-poised if, in addition,
$a_1 = -a_2 = qb_1 = -qb_2$. Further, the series ${}_k\psi_k$ is called
balanced if $b_1\cdots b_k = a_1\cdots a_kq^2$ and $z = q$.
For the most general balanced very-well-poised series
${}_8\psi_8$ all six parameters $z,s_1,s_2,s_3,v_1,v_2$ in
\eqref{psi88} should be arbitrary.}
balanced very-well-poised series
$\;_8\psi_8$. The only similar result we were able to find in the
literature is given by Schlosser (see Equation (2.5) in \cite{Sch06}). However,
his formula is different, since it requires completely opposite
conditions on the arguments in \eqref{psi88}.
In our case the parameters $v_1,v_2$ are
arbitrary indeterminates, but $s_1, s_2, s_3$ and $(-z)$ are restricted
to integral powers of $q$. In Schlosser's case the parameters
$v_1^2, v_2^2$ are restricted to integral powers of $q$, but
$s_1, s_2, s_3, z$ are arbitrary.

Finally note, that for  positive values of $v_1$, $v_2$ and $q<1$, such that
\beq
q^\hf<v_1<1,\qquad q^\hf<v_2<1,\qquad q^\hf<v_1v_2<1\ ,
\eeq
all terms of the series \eqref{hyper} are positive for arbitrary
$ \sigma_1,\sigma_2,\sigma_3,\zeta\in{\mathbb Z}$ (and the series is,
of course, convergent). That is what makes this series important for
applications in statistical mechanics.


\newcommand\oneletter[1]{#1}\def\cprime{$'$}

\end{document}